\documentclass[10pt,journal,final]{IEEEtran}
\usepackage{amsmath,amsfonts}
\usepackage{amsthm}
\usepackage{amssymb}
\usepackage{algorithmic}
\usepackage{algorithm}
\usepackage{array}
\usepackage{mathrsfs}
\usepackage[caption=false,font=footnotesize,labelfont=rm,textfont=rm]{subfig}
\usepackage{textcomp}
\usepackage{stfloats}
\usepackage{url}
\usepackage{amsmath}
\usepackage{mathtools}
\usepackage{verbatim}
\usepackage{graphicx}
\usepackage{cite}
\usepackage{autobreak}
\usepackage{nicematrix}
\usepackage{arydshln}
\usepackage{latexsym}
\usepackage{booktabs}
\usepackage{nomencl}

\makenomenclature
\allowdisplaybreaks[4]

\usepackage{etoolbox}
\renewcommand\nomgroup[1]{%
	\item[\bfseries
	\ifstrequal{#1}{A}{Abbreviations}{%
	\ifstrequal{#1}{B}{Indices and Sets}{%
	\ifstrequal{#1}{C}{Parameters}{%
	\ifstrequal{#1}{D}{Variables}{%
	\ifstrequal{#1}{E}{Functions}{}}}}}%
]}

\begin{document}
\title{Synergising Hierarchical Data Centers and Power Networks: A Privacy-Preserving Approach}

\author{Junhong~Liu,~\IEEEmembership{Member,~IEEE},~Fei~Teng,~\IEEEmembership{Senior~Member,~IEEE},~Yunhe~Hou,~\IEEEmembership{Senior~Member,~IEEE}
        % <-this % stops a space
%\thanks{This work was supported in part by  .}
\thanks{This work was supported in part by the National Key R\&D Program of China under Grant 2023YFA1011301; in part by the National Natural Science Foundation of China (NSFC) under Grant 52177118 and Grant 52437005; and in part by the Research Grants Council of Hong Kong under Grant GRF 17201524. \textit{(Corresponding author: Yunhe Hou.)}}

\thanks{Junhong Liu and Yunhe Hou are with the Department of Electrical and Electronic Engineering, The University of Hong Kong, Hong Kong SAR, China, and also with the Shenzhen Institute of Research and Innovation, The University of Hong Kong, Shenzhen, 518063, China (e-mail: jhliu@eee.hku.hk; yhhou@eee.hku.hk).}

\thanks{Fei Teng is with the Department of Electrical and Electronic Engineering, Imperial College London, London, UK (e-mail: f.teng@imperial.ac.uk).} \vspace{-1.5em}}

% The paper headers
\markboth{ }%
{Shell \MakeLowercase{\textit{et al.}}: A Sample Article Using IEEEtran.cls for IEEE Journals}
\maketitle

%abstract%
\begin{abstract}
In the era of digitization, data centers have emerged as integral contributors sustaining our interlinked world, bearing responsibility for an increasing proportion of the world's energy consumption. To facilitate the their fast rollout while progressing towards net-zero energy systems, the synergy of hierarchical data centers (cloud-fog-edge) and power networks can play a pivotal role. However, existing centralized co-dispatch manners encroach on the privacy of different agents within the integrated systems, meanwhile suffering from the combinatorial explosion. In this research, we propose a near-optimal distributed privacy-preserving approach to solve the non-convex synergy (day-ahead co-dispatch) problem. The synergy problem is formulated as a mixed integer quadratically constrained quadratic programming considering both communication and energy conservation, where Lyapunov optimization is introduced to balance operating costs and uncertain communication delays. To mitigate impacts of the highly non-convex nature, the normalized multi-parametric disaggregation technique is leveraged to reformulate the problem into a mixed integer non-linear programming. To further overcome non-smoothness of the reformulated problem, the customized $\ell_1-$surrogate Lagrangian relaxation method with convergence guarantees is proposed to solve the problem in a distributed privacy-preserving manner. {The effectiveness, optimality, and scalability of the proposed methodologies for the synergy problem are validated via numerical simulations. The results also indicate that computing tasks can be migrated within the hierarchical data centers, demonstrating the flexible resource allocation capabilities of the hierarchical architecture, further facilitating peak load balancing in the power network.}
\end{abstract}


{
% makeindex co-dispatch.nlo -s nomencl.ist -o co-dispatch.nls
\mbox{}
%Abbreviations
\nomenclature[A]{\(cha/dis\)}{Charge/discharge}
\nomenclature[A]{\(iot,fdc,cdc\)}{Internet of things (IoT), fog data center, cloud data center}
\nomenclature[A]{\(ess\)}{Energy storage system}
\nomenclature[A]{\(tran\)}{Transimission (for data flows)}
\nomenclature[A]{\(cal\)}{Calculation}
\nomenclature[A]{\(inf\)}{Infinity}
\nomenclature[A]{\(sup\)}{Supremum}
%Indices and Sets
\nomenclature[B]{\(i,i^{'},j,m\)}{Index of nodes/lines/agents}
\nomenclature[B]{\(\tau\)}{Index of time slot}
\nomenclature[B]{\(k\)}{Index of iterations}
\nomenclature[B]{\(e\)}{Index of segments in RNMDT}
\nomenclature[B]{\(l,u\)}{Index of lower/upper bounds for segments in RNMDT}
\nomenclature[B]{\(p(d)\)}{Index of nodes with power injections (data centers)}
\nomenclature[B]{\({\cal N}_g\)}{Set of traditional thermal generators}
\nomenclature[B]{\({\cal N}_{ess}\)}{Set of energy storage systems}
\nomenclature[B]{\({\cal N}_{iot/fdc/cdc}\)}{Set of IoT/fog/cloud data centers}
\nomenclature[B]{\({\cal C}_{j}\)}{Set of child nodes for node $j$}
\nomenclature[B]{\({\cal A}_{j}\)}{Set of parent nodes for node $j$}
\nomenclature[B]{\({\cal N}_{p}\)}{Set of nodes with power injections}
\nomenclature[B]{\({\cal N}_{b}\)}{Set of branch lines}
\nomenclature[B]{\({\cal N}\)}{Set of total nodes}
\nomenclature[B]{\({\cal D}\)}{Set of total agents in the distributed optimization}
\nomenclature[B]{\({\mathbb Z}^{-}\)}{Set of the negative integers}

%Parameters
\nomenclature[C]{\(p_{j,\tau}^r\)}{Active power for the renewable generator at the node $j$ at time $\tau$}
\nomenclature[C]{\(\pi_{\tau}\)}{Price of charging/discharging per unit power for generators/ess}
\nomenclature[C]{\(\kappa_{\tau}\)}{Price of generating per unit power for generators}
\nomenclature[C]{\({\cal T}\)}{Total number of time slots}
\nomenclature[C]{\(p_{j,\tau}^{\min}/p_{j,\tau}^{\max}\)}{Lower/Upper bound of nodal active power injection of node $j$ at time $\tau$}
\nomenclature[C]{\(q_{j,\tau}^{\min}/q_{j,\tau}^{\max}\)}{Lower/Upper bound of nodal reactive power injection of node $j$ at time $\tau$}
\nomenclature[C]{\(v_{j,\tau}^{\min}/v_{j,\tau}^{\max}\)}{Lower/Upper bound of squared voltage magnitude of node $j$ at time $\tau$}
\nomenclature[C]{\(P_{ij,\tau}^{\min}/P_{ij,\tau}^{\max}\)}{Lower/Upper bound of active power flow on the line $ij$ at time $\tau$}
\nomenclature[C]{\(Q_{ij,\tau}^{\min}/Q_{ij,\tau}^{\max}\)}{Lower/Upper bound of reactive power flow on the line $ij$ at time $\tau$}
\nomenclature[C]{\(l_{ij,\tau}^{\min}/l_{ij,\tau}^{\max}\)}{Lower/Upper bound of power current on the line $ij$ at time $\tau$}
\nomenclature[C]{\(R_{j,\tau}^{down}/R_{j,\tau}^{up}\)}{Limits for ramping down/up for generator $j$ at time $\tau$}
\nomenclature[C]{\(\overline{p_{j,\tau}^{dis}}/\overline{p_{j,\tau}^{cha}}\)}{Upper bound for discharging/charging of ess at the node $j$ at time $\tau$}
\nomenclature[C]{\(r_{ij}/x_{ij}/z_{ij}\)}{Resistance/Reactance/Impedance of the line $ij$}
\nomenclature[C]{\(\eta_{cha}/\eta_{dis}\)}{Charging/Discharging coefficient for the ess, i.e., 0.95, 0.95}

\nomenclature[C]{\(H_{j,max}^{iot/fog}\)}{Maximum volumes of data queues for IoT/fog data centers, i.e., 10, 10 Mb/ms}

%for the data centers%
\nomenclature[C]{\(k^{iot}/k^{fdc}\)}{Capacitance switching coefficient determined by the chip architecture of IoT/fog data centers, typically $1.7 \times 10^{-27}, 1.6 \times 10^{-27} W*s^3/cycle^3$}
\nomenclature[C]{\(d^{iot}/d^{fdc}\)}{Calculation density of IoT/fog data centers, where processing one bit of the data needs $d$ CPU cycle frequencies, typically $5000 cycle/bit$}
\nomenclature[C]{\(f_{j,\tau}^{iot/fdc,min}/f_{j,\tau}^{iot/fdc,max}\)}{Lower/Upper bound of CPU operating frequency of IoT/fog data center $j$ at time $\tau$, i.e., $0, 0, 5\times10^9, 5\times10^9 cycle/s$}
\nomenclature[C]{\(\xi^{iot}/\xi^{fdc}\)}{Time slot length for the IoT/fog data centers, i.e., 1, 1 ms}
\nomenclature[C]{\(g_{j,\tau}^{iot,tran}/g_{j,\tau}^{fdc,tran}\)}{Transmitting power from IoT to fog data centers/from fog to cloud data centers, typically uniform distribution $U(0,1)\times2$ W}
\nomenclature[C]{\(R_{j,\tau}^{iot,tran}/R_{j,\tau}^{fdc,tran}\)}{Data transmission rate from IoT to fog data centers/from fog to cloud data centers}
\nomenclature[C]{\({\cal W}_{j,\tau}^{iot,tran}/{\cal W}_{j,\tau}^{fdc,tran}\)}{Channel bandwidth from IoT to fog data centers/from fog to cloud data centers, typically 40MHz}
\nomenclature[C]{\(\sigma_{iot,tran}^2/\sigma_{fdc,tran}^2\)}{Noise power from IoT to fog data centers/from fog to cloud data centers, typically $-50\sim-30$ dBm}
\nomenclature[C]{\(h_{j,\tau}^{iot,tran}/h_{j,\tau}^{fdc,tran}\)}{Channel gain from IoT to fog data centers/from fog to cloud data centers, typically  $-5\sim-2$}
\nomenclature[C]{\(S_{j,\tau}^{\min}/S_{j,\tau}^{\max}\)}{Lower/Upper bound of state of the charge for the ess $j$ at time $\tau$, i.e., 0, 10 MWh}

\nomenclature[C]{\(I_{j}\)}{Bit amount per task request per time slot arriving at the cloud server $j$, i.e., 2.5 request/Mb}
\nomenclature[C]{\(M_{j, cdc}\)}{Number of servers in cloud data center $j$, i.e., 500}
\nomenclature[C]{\(\tau_{cdc}\)}{Maximum task latency in cloud data centers, i.e., 10s}
\nomenclature[C]{\(P_{peak}/P_{idle}\)}{Peak/Idle power consumption of cloud data centers’ IT equipment, i.e., 500, 200W}
\nomenclature[C]{\(\zeta_j\)}{Power usage effectiveness (PUE), i.e., 1.4}
\nomenclature[C]{\(V_{iot}/V_{fdc}\)}{Hyper-parameters helping balance the economic benefit and stability of queues, i.e., 0.0008, 0.005}
\nomenclature[C]{\(S_{j,\tau}^{iot}\)}{Incoming task requests at IoT data center $j$ at time $\tau$}
\nomenclature[C]{\(H_{j,\tau}^{u}/H_{j,\tau}^{l}\)}{Upper/Lower bound of $H_{j,\tau}^{iot}$}
\nomenclature[C]{\(\triangle u_{j,\tau}\)}{Value of one segment for $H_{j,\tau}^{iot}$}
\nomenclature[C]{\(\triangle w_{j,\tau}\)}{Value of one segment for $H_{j,\tau}^{iot}$ multplied by $U_{j,\tau}^{iot,cal}$}
\nomenclature[C]{\(\vartheta\)}{Index of arbitrary segment in RNMDT}
\nomenclature[C]{\(k_1,k_2\)}{Lower/Upper bound of segments for the integer variable in RNMDT}
\nomenclature[C]{\(\varpi_{i,p},\varpi_{i',p}\)}{Mapping matrices of energy coupling constraints for the agent $i/i'$}
\nomenclature[C]{\(\varpi_{i,d},\varpi_{i',d}\)}{Mapping matrices of communication coupling constraints for the agent $i/i'$}
\nomenclature[C]{\(\eta_p/\eta_d\)}{Penalty factors for energy/communication coupling constraints, i.e., initially, 80, 200}
\nomenclature[C]{\(\boldsymbol{\eta}_{p(d)}^{k}\)}{Vector of penalty factors for the energy/communication coupling constraints at iteration $k$}

\nomenclature[C]{\(\xi^{k}\)}{Stepsize for distributed update at iteration $k$}
\nomenclature[C]{\(\alpha_{k}/\theta/r/c\)}{Parameters for the Polyak step-size rule}
\nomenclature[C]{\(w\)}{Parameter for adaptively adjusting the pernalty terms, i.e., 1.01}
\nomenclature[C]{\(\beta\)}{Parameter for reaching an approximate linear rate}
\nomenclature[D]{\(B_{iot}/B_{fdc}\)}{Constant part of the Lyapunov drift for the queue at IoT/fog data center}
\nomenclature[D]{\(f_{j,\tau}^{iot}/f_{j,\tau}^{fdc}\)}{CPU operating frequency of IoT/fog data centers $j$ at time $\tau$}
%Variables

\nomenclature[D]{\(z_{j, \tau}^{cha}\)}{Charging state of ess $j$ at time $\tau$}
\nomenclature[D]{\(S_{j, \tau}^{ess}\)}{State of charge for ess $j$ at time $\tau$}
\nomenclature[D]{\(p_{j,\tau}/q_{j,\tau}\)}{Active/Reactive power injection of node $j$ at time $\tau$}
\nomenclature[D]{\(v_{j,\tau}\)}{Squared voltage magnitude of node $j$ at time $\tau$}
\nomenclature[D]{\(P_{ij,\tau}/Q_{ij,\tau}/l_{ij,\tau}\)}{Active power/Reactive power/Current flow over the line $ij$ at time $\tau$}
\nomenclature[D]{\(p_{j,\tau}^g/p_{j,\tau}^{dis}/p_{j,\tau}^{cha}\)}{Active power for the generator/discharging of ess/charging of ess at the node $j$ at time $\tau$}
\nomenclature[D]{\(p_{j,\tau}^{iot/fdc/cdc}\)}{Active power for the IoT/fog/cloud data centers at the node $j$ at time $\tau$}

\nomenclature[D]{\(U_{j,\tau}^{iot,cal}/U_{j,\tau}^{iot,tran}\)}{Bit amount of tasks calculated at/transmitted from IoT data center $j$ at time $\tau$}
\nomenclature[D]{\(U_{j,\tau}^{fdc,cal}/U_{j,\tau}^{fdc,tran}\)}{Bit amount of tasks calculated at/transmitted from fog data center $j$ at time $\tau$}
\nomenclature[D]{\(p_{j,\tau}^{iot,cal}/p_{j,\tau}^{iot,tran}\)}{Power consumed for local task computation/transmission of IoT data center $j$ at time $\tau$}
\nomenclature[D]{\(p_{j,\tau}^{fdc,cal}/p_{j,\tau}^{fdc,tran}\)}{Power consumed for local task computation/transmission of fog data center $j$ at time $\tau$}
\nomenclature[D]{\(H_{j,\tau}^{iot}/H_{j,\tau}^{fdc}\)}{Queue of tasks at the IoT/fog data center $j$ at time $\tau$}
%\nomenclature[D]{\(p_{j,\tau}^{iot}/p_{j,\tau}^{fdc}\)}{Total Power consumed for task processing of IoT/fog data center $j$ at time $\tau$}
\nomenclature[D]{\(n_{j, \tau}/\mu_{j,\tau}/\lambda_{j, \tau}\)}{Number of working servers/Utilization rate/Total income bit amount of tasks at the cloud data center $j$ at time $\tau$}
\nomenclature[D]{\(\widehat{{\cal C}}_{j}^{iot}/\widehat{{\cal C}}_{j}^{fdc}\)}{Total approximate cost for the IoT/fog data center $j$ at time $\tau$}
\nomenclature[D]{\({\cal C}_{j}^{iot}/{\cal C}_{j}^{fdc}\)}{Total original cost for the IoT/fog data center $j$ at time $\tau$}
\nomenclature[D]{\(\triangle(H_{j,\tau}^{iot})\)}{Lyapunov drift term for data queue from IoT data center $j$ at time $\tau$}

%RNMDT%
\nomenclature[D]{\(\triangle u_{j,\tau}/\triangle w_{j,\tau}\)}{Convex hull for the segment of one variable/ the segmented multiplication of two variables for IoT data center $j$ at time $\tau$}
\nomenclature[D]{\(z_{j,e,\tau}^{iot}/\hat{y}_{j,e,\tau}\)}{Discrete variable for each segment $e$/Value of the segmented multiplication of two variables for IoT data center $j$ at time $\tau$}
\nomenclature[D]{\(z_{e}^{cdc}/\hat{n}_{e}\)}{Discrete variable for each segment $e$/Value of the segmented multiplication of two variables for cloud data center $j$ at time $\tau$}

%proof%

%distributed optimization%
\nomenclature[D]{\(\boldsymbol{x}_{i,\tau}\)}{Vector of decision variables for each agent $i$ at time $\tau$}
\nomenclature[D]{\(\gamma_p^{k}/\gamma_d^{k}\)}{Primal/Dual residual at iteration $k$}

%Functions
\nomenclature[E]{\(L(H_{j,\tau}^{iot})\)}{Lyapunov function for data queue from IoT data center $j$ at time $\tau$}
\nomenclature[E]{\({\mathcal J}(\centerdot)/{\mathcal J}_{i,\tau}(\centerdot)\)}{Objective function/Agent-based objective function for agent $i$ at time $\tau$ of the reformulated problem by Lyapunov optimization}
\nomenclature[E]{\(\mathcal{P}/\mathcal{P}_{\vartheta}\)}{Original optimzation function/Extended real-value function}
\nomenclature[E]{\(||\centerdot||_1\)}{Norm-1 function for $\centerdot$}

\nomenclature[E]{\({\mathcal L}_{\eta}(\centerdot)/{\mathcal L}_{\eta,i}(\centerdot)\)}{Augmented Lagrangian function/Augmented Lagrangian function for agent $i$ with penalty term $\eta$}

\printnomenclature[2.2cm]
}

\section{Introduction}
% data center (economy, spatial-temporal transition)
% edge-fog-cloud (hiarchial, heteregenous)
% data center%
\IEEEPARstart{D}{ata} centers serve as the foundational infrastructure supporting a myriad of activities across economic and technological domains. Fueled by data-intensive innovations such as artificial intelligence and 5G devices, data centers operate at a high capacity to meet diverse computing requests. Cloud computing is a centralized model where data is stored and processed at a remote data center, while the confluence of data proliferation and long physical distance between cloud data centers and devices can lead to the violation of bandwidth constraints and latency issues \cite{fiandrino2015performance}. To address these challenges, the adoption of Internet of things (IoT) edge and fog data centers has emerged as a prominent solution \cite{ahvar2019estimating}. For instance, Google Inc. launched the Distributed Cloud in 2021, extending its cloud infrastructure to the edge and customer data centers \cite{xu2021cloud}. Edge data centers are strategically positioned closely to end-users, enabling the delivery of rapid edge computing services with minimal latency. Edge computing aims to minimize the volume of data transferred to the cloud data center, thereby reducing the network latency. Moreover, fog data centers co-located with routers in the local network further extend the capabilities of edge computing by providing an intermediary layer of computing infrastructure between edge devices and the cloud data center. Nevertheless, heterogeneity of the hierarchical computing paradigm, i.e., edge-fog-cloud computing, raises multiple challenges, e.g., where to offload the workloads (from the edge or fog servers) and how to meet the quality of service (QoS) requirements \cite{kimovski2021cloud}. {Meanwhile, substantial surge in the digital contents is transforming data centers into one of the fastest-growing consumers of electricity, whose global demand is 299 TWh in 2020 and is estimated to reach 848 TWh in 2030 \cite{mytton2022sources}. Therefore, enhancing energy efficiency of data centers is paramount to reach the carbon neutrality by 2060, requiring coordinated dispatch of the hierarchical data centers and power networks \cite{han2024novel}.}

%eg. however centralized manner distributed manner, privicy issues, memanwhile nonconvex nonsmooth
The prevailing approach for data centers to collaboratively evolve with power networks involves their participation in demand response programs. Researchers have proposed data center-grid coupling models considering queuing theory and job scheduling techniques \cite{zhang2024data}, the data-driven assessment scheme of data centers \cite{cao2022data}, prediction-based pricing with analytic and worst-case bounds \cite{liu2014pricing}, robust bi-level co-optimization model \cite{weng2023distributed}, four-level joint optimal dispatch model \cite{han2024novel} to promote the active participation of data centers in demand response programs. The potential benefits of data centers engaging in demand response services are analyzed in \cite{liu2013data}, with results demonstrating up to a 40\% reduction in data center operational costs. {The aforementioned optimization problems for power networks with data centers are primarily addressed in a centralized manner. However, optimization models for the co-dispatch problem must acknowledge the unavailability of information within data centers to power systems and also the issue of curse of dimension introduced by integer variables\cite{xiao2022privacy,li2023decentralized}. Meanwhile, the overall energy efficiency of hierarchical edge-fog-cloud data centers have rarely been investigated. Addressing these challenges necessitates a systematic solution that respects the interests and privacy of data centers while ensuring overall computational efficiency.}

%current distributed algorithm
{Currently, numerous distributed algorithms have been proposed to solve optimization problems in the integrated energy systems while preserving data privacy, such as the optimal condition decomposition (OCD) \cite{rabiee2012voltage}, alternating direction method of multipliers (ADMM) \cite{erseghe2014distributed}, auxiliary problem principle (APP) \cite{hur2002evaluation}, augmented lagrangian alternating direction inexact newton (ALADIN) \cite{engelmann2018toward}. Among them, ADMM-based approaches have garnered widespread attention for the optimization problems of integrated energy systems due to their strong convergence performance. For instance, ADMM is employed for the the online distributed energy management of data centers \cite{yu2016distributed,zhang2020distributed}. Reference \cite{li2023decentralized} proposed a learning-aided ADMM algorithm for the decentralized optimization for integrated electricity–heat systems with data centers. Nevertheless, these works consider the simplified convex data center models and only focus on the real-time operation of data centers. Moreover, to enhance privacy preservation across data centers, differential privacy techniques have been proposed to obscure sensitive information \cite{zhou2022pgpregel,fan2020privacy}. However, these approaches are typically applied at the individual data center level, overlooking the coordination and synergy required at the system-wide scale.}

%drwabacks
To facilitate the synergy (co-dispatch) of hierarchical data centers and power networks, the optimization problem can be inherently modeled as the mixed integer non-linear programming (MINLP). However, the classic distributed algorithms, i.e., OCD, ADMM, APP, ALADIN, can not handle the non-convex MINLP due to the co-existence of mixed integers and non-linearity \cite{murray2018mixed}. While ALADIN has been reported to perform effectively for the nonlinear programming (NLP), it remains challenging to guarantee the convergence of mixed integer programming (MIP) \cite{murray2018hierarchical}. Beyond the aforementioned distributed algorithms, researchers proposed customized distributed algorithms for solving the optimization problem with integer variables. The $\ell_1$-augmented Lagrangian method (ALM) with additional shares of non-convex reverse norm cuts is proposed for the two-block mixed integer linear programming (MILP) \cite{sun2021decomposition}. Nevertheless, the effectiveness of this method diminishes as the problems scale increases, since it requires the preservation of all historical cuts as constraints. Additionally, SDM-GS-ALM is proposed for solving MILP in a distributed manner \cite{chen2018decentralized, boland2019parallelizable}. {Moreover, to enhance the performance of SDM-GS-ALM, reference \cite{sharma2023novel,sharma2024enhanced} further proposed the VILS algorithm, which adaptively selects inner loop iterations in the computation of Lagrangian upper bounds, rather than relying on pre-fixed inner loop iterations as in SDM-GS-ALM.} However, both SDM-GS-ALM and VILS algorithms require a linear cost function and the construction of a convex hull for integer variables to ensure convergence.

{Nevertheless, several characteristics define the synergy problem of hierarchical data centers and power networks, representing a more general and realistic case in the dispatch/operation of integrated power networks: 1) a potentially large number of integer variables, resulting in the high combinatorial complexity; 2) a non-linear objective function; 3) the presence of discrete variables within the sub-problems. These properties pose obstacles for the applicability of existing distributed algorithms and research. Specifically, the potentially large number of integer variables necessitates sharing numerous non-convex reverse norm cuts for the $\ell_1$-ALM, which diminishes the effectiveness of this method. Meanwhile, the non-linear objective function renders both the SDM-GS-ALM and VILS theoretically inapplicable. Furthermore, the presence of potentially large number of discrete variables within the sub-problems poses significant challenges for employing the ADMM. To overcome these challenges, we propose a set of methodologies inspired by \cite{bragin2018scalable} to solve the large-scale MIQCQP in a privacy-preserving manner. Specifically, to address the non-convex and non-smooth nature of the synergy problem, we propose reformulating the original synergy problem into a MINLP by leveraging the novel reformulated normalized multiparametric disaggregation technique (RNMDT). This reformulation enhances computational efficiency in solving the synergy problem while guaranteeing near-optimality from both theoretical and experimental perspectives. To overcome non-smoothness of the reformulated problem, the customized $\ell_1-$surrogate Lagrangian relaxation method with convergence guarantees is proposed to solve the problem in a distributed privacy-preserving manner. Surrogate subgradient directions are constructed via the surrogate optimality conditions to form acute angles with directions toward the optimal multipliers. With adaptively adjusted stepsizes, multipliers tend to optimal ones, facilitating the attainment of optimal solutions of the non-smooth MINLP problem.} The main contributions are summarized as follows:
\vspace{-0.1cm}

\begin{itemize}
\item{{We formulate the day-ahead co-dispatch problem of hierarchical data center penetrated power networks as a mixed integer quadratically constrained quadratic programming (MIQCQP) considering both communication and energy conservation, where Lyapunov optimization is introduced to minimize operating costs while stabilizing uncertain data queues in the communication network.}}

\item{The highly non-convex synergy problem is reformulated into a mixed integer non-linear programming (MINLP) by leveraging the reformulated normalized multi-parametric disaggregation technique, which can be made arbitrarily precise via a precision parameter.}

\item{{To further overcome combinatorial explosion and preserve privacy for agents, we propose the customized $\ell_1-$surrogate Lagrangian relaxation method with convergence guarantees to solve the non-smooth MINLP in a distributed privacy-preserving manner, which employs surrogate optimality conditions and adaptively adjusted stepsizes to generate search directions that form acute angles toward the optimal multipliers.}}
\end{itemize}
 %\vspace{-0.1cm}
% \vspace{-1.0cm}
The remainder of this paper is organized as follows: Section II introduces the centralized co-dispatch problem of hierarchical data centers integrated power networks and its nearly exact reformulations. Section III proposes a distributed privacy-preserving approach for solving the co-dispatch problem. Section IV demonstrates the effectiveness of the proposed approach with case studies. Section V draws the conclusions.

%power system, themal, ess
%fog, edge, cloud

%lyapunov optimization
%rho-box optimization
%surrograte model based optimization
%theoretical conditions

\section{Formulations of the Edge-Fog-Clouds Penetrated Power Systems}
\subsection{Centralized Co-Dispatch Problem of Hierarchical Data Centers Penetrated Power Networks}
\begin{figure}[!htbp]
	\centering
	\includegraphics[width=1.0\linewidth]{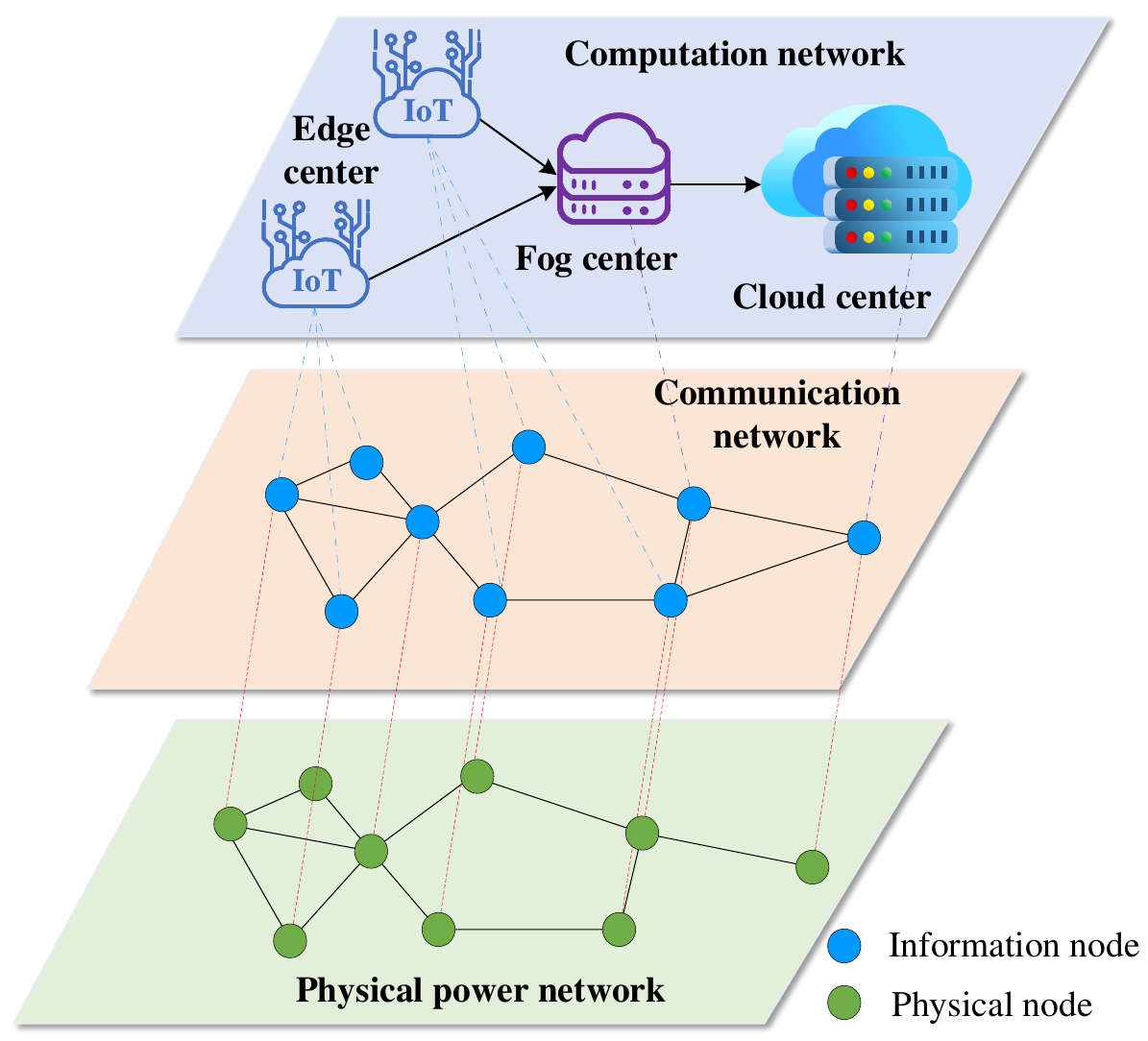}
	\caption{Synergy of the hierarchical data centers and power networks.}
%\vspace{-1.0cm}
	\label{fig1}
\end{figure}

{As illustrated in Fig. \ref{fig1}, this paper considers the day-ahead co-dispatch problem for the hierarchical data centers and distribution power networks \cite{si2020connectivity,li2022edge}, which operate at low to medium voltage levels. In this framework, hierarchical data centers and storage systems are modeled as flexible loads, while distributed generators supply electricity within the distribution system. We formulate the problem as a multi-objective minimization that aggregates economically distinct cost components across stakeholders.} The synergy problem is formulated as:
\begin{subequations}
\begin{align}
&\min\limits_{\boldsymbol{x}}\sum\limits_{\tau \in {\cal T}}\{\sum\limits_{j \in {\cal N}_g} \kappa_{\tau}p_{j,\tau}^{g}+\sum\limits_{j \in {\cal N}_{ess}}\pi_{\tau}(p_{j,\tau}^{cha}-p_{j,\tau}^{dis})+\notag \\
&\sum\limits_{j \in {\cal N}_{cdc}} \pi_{\tau}p_{j,\tau}^{cdc}+\sum\limits_{ij \in {\cal N}_b}  {r_{ij}}l_{ij,\tau}\}+\sum\limits_{j \in {\cal N}_{iot}} {\cal C}_{j}^{iot} +\sum\limits_{j \in {\cal N}_{fdc}} {\cal C}_{j}^{fdc} \label{1a} \\
&\mathbf{s.t.} \sum\limits_{m \in {\cal C}_j}P_{jm,\tau} -\sum\limits_{i \in {\cal A}_j}(P_{ij,\tau}-r_{ij}l_{ij,\tau}) = p_{j,\tau}, j \in {\cal N}_p \label{1b}\\
&\sum\limits_{m \in {\cal C}_j}Q_{jm,\tau} -\sum\limits_{i \in {\cal A}_j}(Q_{ij,\tau}-x_{ij}l_{ij,\tau}) = q_{j,\tau} , j \in {\cal N}_p \label{1c}\\
& v_{j,\tau} = v_{i,\tau}-2({r_{ij}}{P_{ij,\tau}} + {x_{ij}}{Q_{ij,\tau}}) +z_{ij}l_{ij,\tau}, j \in {\cal N}_p  \label{1d} \\
& P_{ij,\tau}^2+Q_{ij,\tau}^2 \le l_{ij,\tau}v_{i,\tau}, ij \in {\cal N}_{b}  \label{1e}\\
&p_{j,\tau}^{\min} \le p_{j,\tau} \le p_{j,\tau}^{\max}, j \in {\cal N}_p \label{1f}\\
&q_{j,\tau}^{\min } \le {q_{j,\tau}} \le q_{j,\tau}^{\max}, j \in {\cal N}_p \label{1g}\\
&v_{j,\tau}^{\min} \le {v_{j,\tau}} \le v_{j,\tau}^{\max}, j \in {\cal N}_p \label{1h}\\
&l_{ij,\tau}^{\min} \le {l_{ij,\tau}} \le l_{ij,\tau}^{\max}, ij \in {\cal N}_{b} \label{1i}\\
&0 \le p_{j,\tau}^{cha} \le z_{j,\tau}^{cha} \overline{p_{j,\tau}^{cha}}, j \in {\cal N}_{ess}  \label{1j}\\
&0 \le p_{j,\tau}^{dis} \le (1-z_{j,\tau}^{cha})\overline{p_{j,\tau}^{dis}}, j \in {\cal N}_{ess}  \label{1k}\\
&S_{j, \tau+1}^{ess}=S_{j, \tau}^{ess}+p_{j, \tau}^{cha}\eta_{cha}-p_{j,\tau}^{dis}/\eta_{dis}, j \in {\cal N}_{ess} \label{1l}\\
&S_{j, \tau}^{min} \le S_{j, \tau}^{ess} \le S_{j, \tau}^{max}, j \in {\cal N}_{ess} \label{1m}\\
& p_{j,\tau}^{g, min} \le  p_{j,\tau}^{g} \le p_{j,\tau}^{g, max}, j \in {\cal N}_g \label{1n}\\
& p_{j,\tau+1}^{g} -p_{j,\tau}^{g} \le R_{j,\tau}^{up}, j \in {\cal N}_g \label{1o}\\
& p_{j,\tau-1}^{g} -p_{j,\tau}^{g} \le R_{j,\tau}^{down}, j \in {\cal N}_g \label{1p}\\
&p_{j,\tau}^{g}+p_{j,\tau}^{r}+p_{j,\tau}^{dis}-p_{j,\tau}^{cha}-p_{j,\tau}^{iot,fdc,cdc}=p_{j,\tau}, j \in {\cal N},  \label{1q}
\end{align}
\end{subequations}
where $z_{j,\tau}^{cha}$ is the binary variable. $\kappa_{\tau}$ is the per unit generation cost, and $\pi_{\tau}$ is the predicted locational marginal price. The objective function \eqref{1a} is to minimize the total operation cost or, equivalently, maximize the social welfare, which contains six parts: i) the first item models the local generation cost; ii) the second item models the expense of charging/discharging electricity by the energy storage system; iii) the third term models the cost of purchasing electricity by the IoT edge data center; iv) the fourth term models the cost of purchasing electricity by the fog data center; v) the fifth term models the cost of purchasing electricity by the cloud data center; vi) the sixth term models the power loss within the system \cite{binetti2014distributed}. {The non-convex constraint in Distflow model is relaxed through the second order cone programming (SOCP) as in \eqref{1e}, which is an exact reformulation for radial networks when power loss minimization is considered \cite{gan2012exact, chowdhury2023socp}.} As denoted in \eqref{1q}, active power injections at each bus are modeled to ensure energy conservation depending on the existence of the generator, renewable generator, energy storage system, IoT edge data center, fog data center, and cloud data center. 

%\vspace{-0.1cm}

\begin{figure}[!htbp]
	\centering
	\includegraphics[width=3.5in]{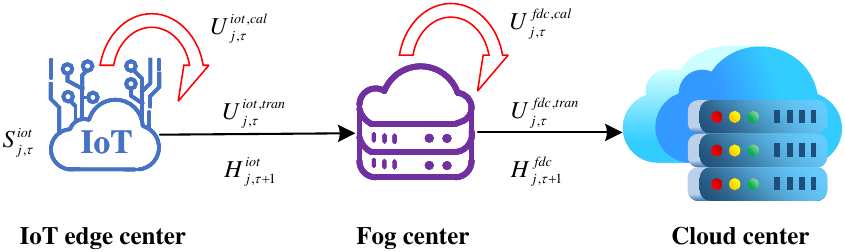}
	\caption{Task allocation of the hierarchical edge-fog-cloud computing.}
	%\vspace{-0.5cm}
	\label{fig2}
\end{figure}

%fog center
According to \cite{ahvar2019estimating},  the three-tier hierarchical data centers, i.e., IoT edge–fog–cloud data centers, can be constructed as in Fig. \ref{fig2}. Power consumed at the IoT edge data center, i.e., $p_{j,\tau}^{fdc}$, consists of two parts: i) the power consumed for local computation at edge data center, i.e., $p_{j,\tau}^{iot,cal}$; ii) power consumed for task transmission from IoT to fog data centers, i.e., $p_{j,\tau}^{iot,tran}$. Meanwhile, power consumed at the fog data center also consists of two parts: i) power consumed for local computation at the fog data center, i.e., $p_{j,\tau}^{fdc,cal}$; ii) and power consumed for task transmission from fog to cloud data centers, i.e., $p_{j,\tau}^{fdc,tran}$. The power consumption for IoT edge and fog data centers can be modelled as:
\begin{subequations}
	\begin{align}
	&p_{j,\tau}^{iot,cal}=k^{iot}U_{j,\tau}^{iot,cal}d^{iot}(f_{j,\tau}^{iot})^2  \label{2a}\\
	&f_{j,\tau}^{iot,min} \le f_{j,\tau}^{iot} \le f_{j,\tau}^{iot,max}  \label{2b}\\
	&0 \le U_{j,\tau}^{iot,cal} \le  \frac{f_{j,\tau}^{iot,max}\xi^{iot}}{d^{iot}}  \label{2c}\\
    &p_{j,\tau}^{iot,tran}=\frac{U_{j,\tau}^{iot,tran}g_{j,\tau}^{iot,tran}}{R_{j,\tau}^{iot,tran}}  \label{2d}\\
    &R_{j,\tau}^{iot,tran}={\cal W}_{j,\tau}^{iot,tran} log_2(1+\frac{h_{j,\tau}^{iot,tran}g_{j,\tau}^{iot,tran}}{\sigma_{iot,tran}^2}) \label{2e}\\
    &0 \le U_{j,\tau}^{iot,tran} \le R_{j,\tau}^{iot,tran}\xi^{iot}  \label{2f}\\
    &H_{j,\tau+1}^{iot}=H_{j,\tau}^{iot}-U_{j,\tau}^{iot,cal}-U_{j,\tau}^{iot,tran}+S_{j,\tau}^{iot} \label{2g}\\ 
    &0 \le U_{j,\tau}^{iot,cal}+U_{j,\tau}^{iot,tran} \le H_{j,\tau}^{iot} \label{2h}\\
    & \lim\limits_{{\cal T} \rightarrow + \infty}  \frac{{\mathbb E} (H_{j,\tau}^{iot}) }{{\cal T}}=0 \label{2i}\\
    &p_{j,\tau}^{fdc,cal}=k^{fdc}U_{j,\tau}^{fdc,cal}d^{fdc}(f_{j,\tau}^{fdc})^2  \label{2j}\\
   &f_{j,\tau}^{fdc,min} \le f_{j,\tau}^{fdc} \le f_{j,\tau}^{fdc,max} \label{2k} \\
   &0 \le U_{j,\tau}^{fdc,cal} \le  \frac{f_{j,\tau}^{fdc,max}\xi^{fdc}}{d^{fdc}} \label{2l}\\
   &p_{j,\tau}^{fdc,tran}=\frac{U_{j,\tau}^{fdc,tran}g_{j,\tau}^{fdc,tran}}{R_{j,\tau}^{fdc,tran}}  \label{2m}\\
   &R_{j,\tau}^{fdc,tran}={\cal W}_{j,\tau}^{fdc,tran} log_2(1+\frac{h_{j,\tau}^{fdc,tran}g_{j,\tau}^{fdc,tran}}{\sigma_{fdc,tran}^2}) \label{2n}\\
   &0 \le U_{j,\tau}^{fdc,tran} \le R_{j,\tau}^{fdc,tran}\xi^{fdc} \label{2o}\\
   &H_{j,\tau+1}^{fdc}=H_{j,\tau}^{fdc}-U_{j,\tau}^{fdc,cal}-U_{j,\tau}^{fdc,tran}+U_{j,\tau}^{iot,tran} \label{2p}\\  
   &0 \le U_{j,\tau}^{fdc,cal}+U_{j,\tau}^{fdc,tran} \le H_{j,\tau}^{fdc} \label{2q}\\
   & \lim\limits_{{\cal T} \rightarrow + \infty}  \frac{{\mathbb E} (H_{j,\tau}^{fdc}) }{{\cal T}}=0 \label{2r}\\
   &p_{j,\tau}^{iot}=p_{j,\tau}^{iot,cal}+p_{j,\tau}^{iot,tran} \label{2s}\\
   & {\cal C}_{j}^{iot}= \lim\limits_{{\cal T} \rightarrow + \infty}  \frac{1}{{\cal T}}\sum_{\tau \in {\cal T}} {\mathbb E} (\pi_{\tau}p_{j,\tau}^{iot}) \label{2t} \\
   &p_{j,\tau}^{fdc}=p_{j,\tau}^{fdc,cal}+p_{j,\tau}^{fdc,tran}  \label{2x}\\
   & {\cal C}_{j}^{fdc}= \lim\limits_{{\cal T} \rightarrow + \infty}  \frac{1}{{\cal T}}\sum_{\tau \in {\cal T}} {\mathbb E} (\pi_{\tau}p_{j,\tau}^{fdc}) \label{2y} \\
   &H_{j,max}^{iot}>= H_{j,\tau}^{iot}>=0, H_{j,max}^{fdc}>=H_{j,\tau}^{fdc}>=0, \label{2g2}
	\end{align}
\end{subequations}
%data center
where \eqref{2f} and \eqref{2o} model the amount of data that can be transferred via the communication channel between IoT edge and fog data centers, and the communication channel between fog and cloud data centers, respectively. Constraints \eqref{2d} and \eqref{2m} model the power consumption of transferring data for IoT edge and fog data centers, which depend on the data transmission rates, i.e., $R_{j,\tau}^{iot,tran}$ and $R_{j,\tau}^{fdc,tran}$ \cite{abbasi2021intelligent}. {For the data transmission, the wireless communication is considered due to the emergence of 5G/6G technologies in the distribution power network \cite{chowdhury20206g}. The Shannon-Hartley formula captures the relation between energy consumption in wireless uplink channels and the data transmission rate, as denoted by formulas \eqref{2e} and \eqref{2n}. This formula has been widely adopted in existing research \cite{chen2018computation,lin2020distributed}. Formulas \eqref{2a} and \eqref{2j} model the energy consumption for processing tasks at the IoT and fog data centers, respectively. Compared to the energy consumed for processing tasks at local data centers, the power required for transmitting data to remote data centers is much lower. However, both types of energy consumption are non-negligible. Formulas \eqref{2g} and \eqref{2p} model the data queues from the IoT to fog data centers and that from the fog to cloud data centers, respectively.} $H_{j,\tau}^{iot}$ and $H_{j,\tau}^{fdc}$ are the data buffer queues exist in the two communication channels, which follows first-in-first-out rule. Due to the potentially uncertain communication demands within each dispatch time-slot, i.e., one hour, the expected costs for IoT edge and fog data centers are employed as in \eqref{2t} and \eqref{2y}. To avoid large deviation from the expected state, the real-time optimization method can also be considered within each dispatch time-slot during the day-ahead dispatch \cite{zhang2021multi}.

For the cloud data center, approximately 40\% of energy consumption comes from cooling systems and the remaining consumption is from the information technology (IT) facilities. {Normally the energy consumption of IT facilities is modeled and an additional coefficient, i.e., power usage effectiveness (PUE) denoted by $\zeta_j$, is multiplied to derive the overall energy consumption \cite{liu2020energy}. More detailed energy consumption models for cloud data centers are also possible \cite{haywood2010sustainable}, including the IT computer rack power, power delivery loss from the power distribution units (PDUs) to the racks, the power for lights, and the power consumed to operate the cooling system. However, incorporating such models introduces additional non-linearity into the synergy problem, posing significant challenges for solving the overall optimization problem.} Meanwhile, the average time delay of task requests at cloud data center follows the M/M/1 queue theory \cite {alshahrani2014modeling}. The modeling of cloud data center follows as:
\begin{subequations}
	\begin{align}
		& \lambda_{j, \tau}/I_{j} \le n_{j,\tau} \le M_{j,cdc}, n_{j,\tau} \in {\mathbb Z}^{+} \label{3a} \\
		& 0 \le \mu_{j,\tau} \le 1 \label{3b}\\
	    &\lambda_{j,\tau}=\sum_{i \in {\cal N}_{fdc}} U_{i,\tau}^{fdc,tran} \label{3c}\\
		& 0 \le \frac{1}{\mu_{j,\tau}-\frac{\lambda_{j,\tau}}{n_{j,\tau}I_{j}}} \le \tau_{cdc} \label{3d}\\
	   & p_{j,\tau}^{cdc} = [(P_{peak}-P_{idle})\mu_{j,\tau}n_{j,\tau}+M_{j,cdc}P_{idle}]\cdot \zeta_j, \label{3e}
	\end{align}
\end{subequations}
where $n_{j,\tau}$ is the number of servers in working conditions and $\mu_{j,\tau}$ is the server utilization rate. $\lambda_{j,\tau}$ is the total incoming computing request at time slot $\tau$ of cloud data center $j$.
%Lyapunov

%\vspace{-0.5cm}

\subsection{Lyapunov Optimization based Corrective Strategy}
{For the online corrective strategy, Lyapunov optimization method exhibits superior performance and lower computational complexity compared to other online methods, e.g., online convex optimization and model predictive control \cite{wang2023online}. Therefore, the real-time online optimization method, i.e., Lyapunov optimization method, is introduced to reformulate the synergy problem with a three-tier data center structure.} Accordingly, limit operations in the objective function as well as constraints \eqref{2i} and \eqref{2r} can be avoided following the method, enabling the problem to be directly handled by commercial solvers. The Lyapunov function and the Lyapunov drift \cite{li2019lyapunov} are employed to define the stability of data queues as:
\begin{subequations}
	\begin{align}
		&L(H_{j,\tau}^{iot}):=\frac{1}{2} (H_{j,\tau}^{iot})^2 \\
		&\triangle(H_{j,\tau}^{iot}):=L(H_{j,\tau+1}^{iot})-L(H_{j,\tau}^{iot}).
	\end{align}
\end{subequations}
{\bf Lemma 1:} The Lyapunov drift term for data queue from IoT edge data center, i.e., $\triangle_{H_{j,\tau}^{iot}}$, can be upper bounded as follow:
\begin{align}
		&\triangle(H_{j,\tau}^{iot}) \le B_{iot}+H_{j,\tau}^{iot}[ S_{j,\tau}^{iot}-U_{j,\tau}^{iot,cal}-U_{j,\tau}^{iot,tran}],
\end{align}
where $B_{iot}$ is a constant value. 

%\sum_{k\in{\cal T}_{fdc}} 
\textit{Proof}: The following relation can be derived according to the definition of Lyapunov drift:
\begin{align}
		\triangle(H_{j,\tau}^{iot})&=\frac{1}{2} (H_{j,\tau+1}^{iot})^2-\frac{1}{2} (H_{j,\tau}^{iot})^2 \notag\\
		&=\frac{1}{2}(S_{j,\tau}^{iot}-U_{j,\tau}^{iot,cal}-U_{j,\tau}^{iot,tran})^2 \\
		&\quad +H_{j,\tau}^{iot}[ S_{j,\tau}^{iot}-U_{j,\tau}^{iot,cal}-U_{j,\tau}^{iot,tran}] \notag\\
		&\le B_{iot}+H_{j,\tau}^{iot}[ S_{j,\tau}^{iot}-U_{j,\tau}^{iot,cal}-U_{j,\tau}^{iot,tran}], \notag
\end{align}
where $B_{iot}$ represents the upper bound for the term $\frac{1}{2}(S_{j,\tau}^{iot}-U_{j,\tau}^{iot,cal}-U_{j,\tau}^{iot,tran})^2$. Similarly, we can obtain the upper bound of the Lyapunov drift for the data queque from the fog data center to cloud data center as:
\begin{align}
		\triangle(H_{j,\tau}^{fdc}) & \le B_{fdc}  \notag\\
		&+ H_{j,\tau}^{fdc}[U_{j,\tau}^{iot,tran}-U_{j,\tau}^{fdc,cal}-U_{j,\tau}^{fdc,tran}].
\end{align}

Based on the obtained upper bounds, the objective function for IoT edge and fog data centers can be reformulated as follows:
\begin{subequations}
	\begin{align}
		\widehat{{\cal C}}_{j}^{iot}= &\sum_{\tau \in {\cal T}} \{\pi_{\tau}p_{j,\tau}^{iot}+\frac{H_{j,\tau}^{iot}}{V_{iot}}[ S_{j,\tau}^{iot}-U_{j,\tau}^{iot,cal}-U_{j,\tau}^{iot,tran}]\} \label{7a} \\
	    \widehat{{\cal C}}_{j}^{fdc}= &\sum_{\tau \in {\cal T}} \{\pi_{\tau}p_{j,\tau}^{fdc}\notag\\
	    &+\frac{H_{j,\tau}^{fdc}}{V_{fdc}}[U_{j,\tau}^{iot,tran}-U_{j,\tau}^{fdc,cal}-U_{j,\tau}^{fdc,tran}] \}, \label{7b}
	\end{align}
\end{subequations}
where $V_{iot}$ and $V_{fdc}$ are hyper-parameters helping balance the economic benefit and stability of queues. As the Lyapunov drift term becomes larger, the queue stability worsens. The objective function in \eqref{2t} and \eqref{2y} can be reformulated as \eqref{7a} and \eqref{7b}. Since the second terms in \eqref{7a} and \eqref{7b} are bounded, under the assumption that optimal W-only policy exists, limiting constraints, i.e. \eqref{2i} and \eqref{2r}, can be automatically guaranteed according to the Lyapunov optimization theory \cite{urgaonkar2009opportunistic}. In this way, these limiting constraints are avoided. {Since the reformulated objective function for the fog center as in \eqref{7a} and \eqref{7b}, and time delay constraints for the M/M/1 queue as in \eqref{3d} contain bi-linear terms, the centralized synergy problem is formulated as a mixed integer quadratically constrained quadratic programming (MIQCQP) as:}
\begin{subequations}
	{
	\begin{align}
		&\min\limits_{\boldsymbol{x}} \sum\limits_{\tau \in {\cal T}} \{\sum\limits_{j \in {\cal N}_g} \kappa_{\tau}p_{j,\tau}^{g}+\sum\limits_{j \in {\cal N}_{ess}}\pi_{\tau}(p_{j,\tau}^{cha}-p_{j,\tau}^{dis})+   \notag \\
		&\sum\limits_{j \in {\cal N}_{cdc}} \pi_{\tau}p_{j,\tau}^{cdc}+\sum\limits_{ij \in {\cal N}_b}  {r_{ij}}l_{ij,\tau}\}+\sum\limits_{j \in {\cal N}_{iot}}\widehat{{\cal C}}_{j}^{iot}+\sum\limits_{j \in {\cal N}_{fdc}} \widehat{{\cal C}}_{j}^{fdc}  \label{c_1} \\
		&\quad  \mathbf{s.t.} \quad  \quad \quad\quad  \eqref{1a}-\eqref{1q}, \eqref{2a}-\eqref{2h},  \\
		& \quad \quad\quad \quad \eqref{2j}-\eqref{2q}, \eqref{2s}-\eqref{2g2}, \eqref{3a}-\eqref{3e}.
	\end{align}
}
\end{subequations}
{The constructed MIQCQP is computationally challenging to solve, even for state-of-the-art commercial solvers \cite{andrade2022p}.}

%reformulation to MILP
\subsection{Reformulated Normalized Multi-parametric Disaggregation Technique}
To overcome heavy computational issues of MIQCQP, we here introduce the novel reformulated normalized multiparametric disaggregation technique (RNMDT) to transform the original non-convex problem into mixed integer non-linear programming (MINLP) at an arbitrary precision, which is crucial to ensure the convergence for the Lagrangian decomposition-based distributed optimization \cite{andrade2022p}. Take one term from \eqref{7a}, i.e., $H_{j,\tau}^{iot}U_{j,\tau}^{iot,cal}$, for instance. It can be reformulated as follows:
\begin{subequations}
	\begin{align}
		&H_{j,\tau}^{iot}U_{j,\tau}^{iot,cal} =U_{j,\tau}^{iot,cal}H_{j,\tau}^{l} \notag \\
		&\quad\quad +(H_{j,\tau}^{u}-H_{j,\tau}^{l})(\sum_{e 
		\in {[\vartheta,-1]}}2^e \hat{y}_{j,e,\tau}+\triangle w_{j,\tau}) \label{8a}\\
		&H_{j,\tau}^{iot}=H_{j,\tau}^{l} +(H_{j,\tau}^{u}-H_{j,\tau}^{l})(\sum_{e 
			\in {[\vartheta,-1]}}2^e z_{j,e,\tau}^{iot}+\triangle u_{j,\tau} ) \label{8a2}\\
		&U_{j,\tau}^{l}z_{j,e,\tau}^{iot} \le \hat{y}_{j,e,\tau} \le U_{j,\tau}^{u}z_{j,e,\tau}^{iot}   \label{8b} \\
	   &U_{j,\tau}^{l}(1-z_{j,e,\tau}^{iot}) \le  U_{j,\tau}^{iot,cal}-\hat{y}_{j,e,\tau} \le  U_{j,\tau}^{u}(1-z_{j,e,\tau}^{iot}) \label{8c} \\
	   &2^\vartheta(U_{j,\tau}^{iot,cal}-U_{j,\tau}^{u})+U_{j,\tau}^{u} \triangle u_{j,\tau} \le \triangle w_{j,\tau}  \label{8d}\\
	   &\triangle w_{j,\tau} \le  2^\vartheta(U_{j,\tau}^{iot,cal}-U_{j,\tau}^{l})+U_{j,\tau}^{l}\triangle u_{j,\tau} \label{8e}\\
	   & U_{j,\tau}^{l} \triangle u_{j,\tau} \le \triangle w_{j,\tau} \le  U_{j,\tau}^{u} \triangle u_{j,\tau}  \label{8f}\\
	   & 0 \le \triangle u_{j,\tau} \le  2^\vartheta, \vartheta \in {\mathbb Z}^{-}, \label{8g}
	\end{align} 
\end{subequations}
where $z_{j,e,\tau}^{iot}$ is the binary variable. The reformulation can be made arbitrarily exact through employing a segment parameter $\vartheta$. {The variable $H_{j,\tau}^{iot}$ is discretized into $|\vartheta|$ parts, with the smallest size of $2^\vartheta$. This bisection method for discretizing the linear variable introduces less binary variables to reach the same precision as the classical McCormick envelopes \cite{mitsos2009mccormick}, where the number of binary variables for RNMDT is $|\vartheta|$.} This approach is applied to all the bi-linear terms in \eqref{7a} and \eqref{7b}. The time delay model in (3d) is simplified as:
\begin{align}
	& {n_{j,\tau}I_{j}} \le  (\mu_{j,\tau}{n_{j,\tau}I_{j}} -\lambda_{j,\tau})\tau_{cdc}, \label{3g}
\end{align} 
where $\mu_{j,\tau}$ is a linear variable and $n_{j,\tau}$ is a non-negative integer. The non-linear term, i.e., $\mu_{j,\tau}n_{j,\tau}$, is reformulated as:
\begin{subequations}
\begin{align}
	& \mu_{j,\tau}n_{j,\tau}= \sum_{e \in [1, k_1]}2^{e-1}\hat{n}_{e}+  (M_{j,cdc}-2^{k_2-1}+1)\hat{n}_{e}\label{10a} \\
	& n_{j,\tau}= \sum_{e \in [1, k_1]}2^{e-1}z_{e}^{cdc}+  (M_{j,cdc}-2^{k_2-1}+1)z_{e}^{cdc} \label{10b} \\
    &  k_1=\lfloor log_2(M_{j,cdc}+1)\rfloor \label{10c}\\
	&  k_2=\lceil log_2(M_{j,cdc}+1)\rceil \label{10d}\\
	& 0 \le \hat{n}_{e} \le z_{e}^{cdc}  \label{10e}\\
	& 0 \le \mu_{j,\tau}- \hat{n}_{e} \le 1-z_{e}^{cdc}, \label{10f}
\end{align} 
\end{subequations}
where $z_{e}^{cdc}$ is the binary variable. This reformation for the non-linear term with a non-negative integer is exact.

Let ${\mathcal J} (\boldsymbol{x})$ be the objective function of the reformulated problem by Lyapunov optimization, i.e., MIQCQP. Let $\sigma_{\Omega}(\boldsymbol{x})$ denotes the indicator function for the constraint set ${\Omega}(\boldsymbol{x})$, where $\sigma_{\Omega}(\boldsymbol{x})=0$ if $\boldsymbol{x} \in {\Omega}(\boldsymbol{x})$, and $\sigma_{\Omega}(\boldsymbol{x})=+\infty$, otherwise. The problem (MIQCQP) can be represented by the infimum of an extended real-value function, i.e., $\mathcal{P}$: min $\{{\mathcal J} (\boldsymbol{x})+\sigma_{\Omega}(\boldsymbol{x})\}$. $\mathcal{P}_{\vartheta}$: min $ \{{\mathcal J}_{\vartheta} (\boldsymbol{x})+\sigma_{\Omega_{\vartheta}}(\boldsymbol{x}) \}$ can represent the relaxed form of $\mathcal{P}$ via RNMDT with an arbitrary parameter $\vartheta,\vartheta<0$.

{\bf Lemma 2:} Assume $0 \in \text{int } \sigma_{\Omega}(\boldsymbol{x})$. Then the following optimality condition holds:
\begin{align}
   \mathop{inf}\limits_{\boldsymbol{x}} \text{ }\mathcal{P}_{\vartheta} \xrightarrow[\vartheta \rightarrow -\infty]{} \mathop{inf}\limits_{\boldsymbol{x}} \text{ } \mathcal{P}. \label{12}
\end{align}

\textit{Proof:} There are two cases for the relaxation.

\textit{Case 1:} When one variable in the non-linear term is non-negative integer, the relaxation via RNMDT over the integer variable is exact. This case is employed for the time delay model as in the \eqref{10a}-\eqref{10f}. The segment parameter $\vartheta$ is fixed as $k_1$ as in \eqref{10c}, the Mccormick envelope is simplified to be exact reformulations as in \eqref{10e} and \eqref{10f}. For this case, the optimality condition \eqref{12} holds true for the fixed parameter $\vartheta$.

\textit{Case 2:} When two variables in the non-linear term are both linear, the relaxation of one of the variables via RNMDT can be made arbitrarily exact. This case is for the Lyapunov drift term as in \eqref{8a}-\eqref{8g}. {The linear variable $H_{j,\tau}^{iot}$ as in \eqref{8a}-\eqref{8g} is actually divided into two parts, i.e., an exact part for $\sum_{e \in {[\vartheta,-1]}}2^e z_{j,e,\tau}^{iot}$ and an approximate part for $\triangle u_{j,\tau}$. The error of this relaxation is mainly introduced by the approximate part, i.e., $\triangle u_{j,\tau}$, which is upper bounded by $2^\vartheta$. This indicates that if the segmentation is large enough, i.e., $\vartheta \rightarrow -\infty$, the error introduced by the discretization method approaches zero.} For arbitrary $\boldsymbol{x^{\dagger}} \in \Omega(\boldsymbol{x})$, the following relations can be obtained following \cite{rockafellar2009variational,andrade2022p}:
%\mathop{\text{sup}}\limits_{\varepsilon>0} \mathop{\text{inf}}\limits_{x \in \Lambda_{\varepsilon}(x^{\dagger})}()
\begin{subequations}
	\begin{align}
		& \mathop{e\text{-lim sup}}\limits_{\vartheta \rightarrow -\infty} ({\mathcal J}_{\vartheta} (\boldsymbol{x^{\dagger}})+\sigma_{\Omega_{\vartheta}}(\boldsymbol{x^{\dagger}})) \le {\mathcal J} (\boldsymbol{x^{\dagger}})+\sigma_{\Omega}(\boldsymbol{x^{\dagger}}) \\
	    & {\mathcal J}(\boldsymbol{x^{\dagger}})+\sigma_{\Omega}(\boldsymbol{x^{\dagger}})  \le \mathop{e\text{-lim inf}}\limits_{\vartheta \rightarrow -\infty} ({\mathcal J}_{\vartheta} (\boldsymbol{x^{\dagger}})+\sigma_{\Omega_{\vartheta}}(\boldsymbol{x^{\dagger}})). 
	\end{align} 
\end{subequations}

These relations further imply that a sequence of relaxed problems, $\{\mathcal{P}_{\vartheta}\}_{-\infty}^{\vartheta=-1}$, epi-converges uniformly to the original problem $\mathcal{P}$ when $\vartheta \rightarrow -\infty$ as follows:
%\begin{subequations}
	\begin{align}
     \{{\mathcal J}_{\vartheta} (\boldsymbol{x})+\sigma_{\Omega_{\vartheta}}(\boldsymbol{x}) \}_{-\infty}^{\vartheta=-1} \xrightarrow[\text{epi-converge}]{}  {\mathcal J} (\boldsymbol{x})+\sigma_{\Omega}(\boldsymbol{x}). 
	\end{align}
%\end{subequations} 
Assume that functions ${\mathcal J}_{\vartheta} (\boldsymbol{x})+\sigma_{\Omega_{\vartheta}}(\boldsymbol{x})$ and ${\mathcal J} (\boldsymbol{x})+\sigma_{\Omega}(\boldsymbol{x})$ are both proper and lower semi-continuous, where ${\mathcal J} (\boldsymbol{x})+\sigma_{\Omega}(\boldsymbol{x})$ is bounded below on bounded sets and ${\mathcal J}_{\vartheta} (\boldsymbol{x})+\sigma_{\Omega_{\vartheta}}(\boldsymbol{x})$ is equi-hypercoercive as follow:
\begin{align}
	 \mathop{\text{lim}}\limits_{x\rightarrow \infty} \frac{{\mathcal J}_{\vartheta} (\boldsymbol{x})+\sigma_{\Omega_{\vartheta}}(\boldsymbol{x})}{||\boldsymbol{x}||_1}= + \infty. \label{cond15}
\end{align}
Equation \eqref{cond15} can be satisfied due to the quadratic terms in the objective function ${\mathcal J}_{\vartheta} (\boldsymbol{x})$. Allied with the premise of $0 \in \text{int } \sigma_{\Omega}(\boldsymbol{x})$, the optimality condition \eqref{12} can be obtained \cite{rockafellar1966level}.

%ADMM algortihm

%write the article take the flows%
\section{Privacy-Preserving Distributed Approach}
{The reformulated small-scale MINLP for the co-dispatch of data center penetrated power networks can be solved by the mature commercial solvers. However, this optimization problem is spatially and temporally correlated. When the scalability of the problem increases, the commercial software has the problem of deriving optimal solutions within finite time due to the curse of dimension introduced by integer variables. Meanwhile, the centralized method infringes on the privacy of different entities in the integrated system. To deal with these issues, we propose the distributed privacy-preserving optimization approach to solve the non-convex and non-smooth problem. Meanwhile, it can also help to mitigate the computational burden for large-scale problem by exploiting the property of exponential reduction of complexity through decoupling spatial correlations and coordinating the sharing of a sequence of decision variables for the distributed computation.}

\subsection{General Formulation of the Co-Dispatch Problem}
For easy decomposition, we can rewrite the problem into a compact form as folalows:
\begin{subequations}
	\begin{align}
		 &\min\limits_{\boldsymbol{x}} \quad {\mathcal J} (\boldsymbol{x}) = \sum_{i \in {\mathcal D}}\sum_{\tau \in {\mathcal T}}   {\mathcal J}_{i,\tau} (\boldsymbol{x}_{i,\tau}) \\
	     &\mathbf{s.t.} \quad  \boldsymbol{x}_{i,\tau} \in {\mathcal X}_i,  \forall i \in {\mathcal D}  \\
	     &\varpi_{i,p}\boldsymbol{x}_{i,\tau}=\varpi_{i',p}\boldsymbol{x}_{i',\tau}, \forall i \in {\mathcal N}_p, i' \in {\mathcal D}/{\mathcal N}_p \\
	     &\varpi_{i,d}\boldsymbol{x}_{i,\tau}=\varpi_{i',d}\boldsymbol{x}_{i',\tau}, \forall i \in {\mathcal N}_{iot,(fdc)}, i' \in {\mathcal N}_{fdc,(cdc)},
	\end{align} 
\end{subequations}
where the set ${\mathcal D}$ is the collection of agents in integrated power systems, i.e., ${\mathcal D}={\mathcal N}_p \cup {\mathcal N}_{ess}  \cup {\mathcal N}_{gen} \cup {\mathcal N}_{iot} \cup {\mathcal N}_{fdc} \cup {\mathcal N}_{cdc} $. Vector $\boldsymbol{x}_{i,\tau}$ contains the decision variables for each entity $i$ at each time step $\tau$. ${\mathcal X}_i$ forms the non-convex feasible region for the decision variable, constructed by the mixed integer non-linear constraints \eqref{1b}-\eqref{1q}, \eqref{2a}-\eqref{2g2}, \eqref{3a}-\eqref{3e}, \eqref{8a}-\eqref{8g}, \eqref{10a}-\eqref{10f}. $ \varpi_{i,p}$ and $ \varpi_{i,d}$ are the mapping matrices for the energy and communication coupling constraints:
\begin{subequations}
	\begin{align}
		 &\varpi_{i,p}\boldsymbol{x}_{i,\tau}=(p_{i,\tau}^{g},p_{i,\tau}^{iot},p_{i,\tau}^{fdc},p_{i,\tau}^{cdc},p_{i,\tau}^{dis}-p_{i,\tau}^{cha})^{T}, \forall i \in {\mathcal N}_p  \label{gc1}\\
		&\varpi_{i,d}\boldsymbol{x}_{i,\tau}=(U_{i,\tau}^{iot,tran},U_{i,\tau}^{fdc,tran})^{T}, \forall i \in {\mathcal N}_{iot}/{\mathcal N}_{fdc}. \label{gc2}
	\end{align} 
\end{subequations}
The reformulated optimization problem is a MINLP. To overcome the non-smoothness of the problem, the $\ell_1-$surrogate Lagrangian relaxation method is proposed to solve the problem in a distributed privacy-preserving manner. 
\begin{subequations}
	\begin{align}
		 \min\limits_{\boldsymbol{x}} \quad&{\mathcal L}_{\eta} (\boldsymbol{x}_i,\boldsymbol{x}_{i'},\boldsymbol{\zeta}_{i,p},\boldsymbol{\zeta}_{i,d}) = \sum_{i \in {\mathcal D}}\sum_{\tau \in {\mathcal T}} \{{\mathcal J}_{i,\tau} (\boldsymbol{x}_{i,\tau}) \} \notag \\
		&+\sum_{i \in {\mathcal N}_p, i' \in {\mathcal D}/{\mathcal N}_p}\sum_{\tau \in {\mathcal T}} \{\boldsymbol{\zeta}_{i,\tau,p}(\varpi_{i,p}\boldsymbol{x}_{i,\tau}-\varpi_{i',p}\boldsymbol{x}_{i',\tau}) \notag\\
		&+\eta_p||\varpi_{i,p}\boldsymbol{x}_{i,\tau}-\varpi_{i',p}\boldsymbol{x}_{i',\tau}||_1 \} \notag\\
		&+\sum\limits_{\substack{i \in {\mathcal N}_{iot,(fdc)}\\, i' \in {\mathcal N}_{fdc,(cdc)}}}\sum_{\tau \in {\mathcal T}} \{\boldsymbol{\zeta}_{i,\tau, d}(\varpi_{i,d}\boldsymbol{x}_{i,\tau}-\varpi_{i',d}\boldsymbol{x}_{i',\tau}) \notag\\
		&+\eta_d||\varpi_{i,d}\boldsymbol{x}_{i,\tau}-\varpi_{i',d}\boldsymbol{x}_{i',\tau}||_{1}\} \label{13a}\\
		&\mathbf{s.t.} \quad  \boldsymbol{x}_{i,\tau},  \boldsymbol{x}_{i',\tau}  \in {\mathcal X}_i,  \forall i, i' \in {\mathcal D}, \label{13b}
	\end{align} 
\end{subequations}
where the global equality constraints \eqref{gc1} and \eqref{gc2} are penalized through the $\ell_1-$norm, helping boost convergence of the distributed optimization. To further increase smoothness of the augmented Lagrangian problem, the $\ell_1-$norm can be reformulated through a set of inequality constraints, we refer interesting readers to \cite{bragin2015convergence,bragin2018scalable} for more details.  

\begin{algorithm}[!htbp]
\caption{Customized Distributed $\ell_1-$Surrogate Lagrangian Decomposition Algorithm.}\label{alg1}
\begin{algorithmic}
\STATE 
%\STATE {\textsc{KeyGen}}
\STATE $  \text{Initialize } \boldsymbol{x}_{i,\tau}^0,  \boldsymbol{x}_{i',\tau}^0, \boldsymbol{\zeta}_{i,\tau,p}^0, \boldsymbol{\zeta}_{i,\tau, d}^0 , \eta_p^0, \eta_d^0, K_{max}, \epsilon$
\STATE  \textbf{For} each iteration $k=0, 1,2,\cdots, K_{max}$ \textbf{do}
\STATE  \hspace{0.2cm} \textbf{For} each agent $i, i \in {\mathcal D}$ \textbf{do}
\STATE  \hspace{0.5cm} Receive variables $\boldsymbol{x}_{i',\tau}^{k}$ from other agents
\STATE  \hspace{0.5cm} Solve the decomposed sub-problem in \eqref{13a}-\eqref{13b}
\STATE  \hspace{0.5cm} Check surrogate optimality condition in \eqref{15a} and \eqref{15b}
\STATE  \hspace{0.6cm}  $\textbf{If}$ the surrogate optimality condition is satisfied:
\STATE  \hspace{0.8cm} Update dual variables following \eqref{16a} and \eqref{16b} 
\STATE  \hspace{0.6cm} $\textbf{else}$
\STATE  \hspace{0.8cm} Update dual variables as $\boldsymbol{\zeta}_{i,\tau,p}^{k+1}=\boldsymbol{\zeta}_{i,\tau,p}^{k}$,$\boldsymbol{\zeta}_{i,\tau,d}^{k+1}=\boldsymbol{\zeta}_{i,\tau,d}^{k}$
\STATE  \hspace{0.6cm} $\textbf{end if}$
\STATE  \hspace{0.5cm} Share variables $\boldsymbol{x}_{i,\tau}^{k+1}$ to other agents
\STATE  \hspace{0.2cm} $\textbf{end for}$
\STATE  \hspace{0.2cm} Update penalty terms following \eqref{20a} and \eqref{20b}
\STATE  \hspace{0.2cm} Check residuals defined in \eqref{18} and \eqref{21}
\STATE  $\textbf{end for}$
%\STATE \hspace{0.5cm}$ \theta \gets [(p-1)(q-1)]^{-1} \text{ }mod\text{ } n $
\end{algorithmic}
\end{algorithm}

%\begin{figure}[!htbp]
%\centering
%%\includegraphics[width=3.5in]{enc1.pdf}
%\caption{Secure two-party encrypted computation mechanism.}
%\vspace{-1.5cm}
%\[\label{fig_enc}\]
%\end{figure}

\subsection{Customized Distributed Privacy-Preserving Approach}
We can define the decomposed sub-problem for each entity at each iteration. Take the sub-problem for the power network operator, i.e., $i \in {\cal N}_p$, for instance. It can be formulated as:
\begin{subequations}
	\begin{align}
		&\min\limits_{\boldsymbol{x}_{i,\tau}} \quad{\mathcal L}_{\eta,i} (\boldsymbol{x}_i,\boldsymbol{x}_{i'},\boldsymbol{\zeta}_{i,p},\boldsymbol{\zeta}_{i,d}) =\sum_{\tau \in {\mathcal T}} \{{\mathcal J}_{i,\tau} (\boldsymbol{x}_{i,\tau})\notag \\
		&+\boldsymbol{\zeta}_{i,\tau,p}\varpi_{i,p}\boldsymbol{x}_{i,\tau} +\eta_p||\varpi_{i,p}\boldsymbol{x}_{i,\tau}-\varpi_{i',p}\boldsymbol{x}_{i',\tau}||_1 \} \\
		&\mathbf{s.t.} \quad  \boldsymbol{x}_{i,\tau},  \boldsymbol{x}_{i',\tau}  \in {\mathcal X}_i,  \forall i \in {\mathcal N}_p, i' \in  {\mathcal D}/{\mathcal N}_p,
	\end{align} 
\end{subequations}
where the power network operator is responsible for calculating the optimal power flow and ensuring the energy power conservation. The fourth and fifth terms in \eqref{13a} are not included for the power network, but will take effects for the IoT edge, fog, or cloud data centers, i.e., $i \in {\cal N}_{iot}/{\cal N}_{fog}/{\cal N}_{cdc}$.

To overcome non-smoothness and facilitate convergence of the optimization problem, the surrogate optimality condition can be defined for agent $ i$, $i \in {\cal D}$, at iteration $k$ as follows:
\begin{subequations}
	\begin{align}
		&{\mathcal L}_{\eta}(\boldsymbol{x}_i^{k+1},\boldsymbol{x}_{i'}^{k},\boldsymbol{\zeta}_{i,p}^{k},\boldsymbol{\zeta}_{i,d}^{k})\le{\mathcal L}_{\eta}(\boldsymbol{x}_i^{k},\boldsymbol{x}_{i'}^{k},\boldsymbol{\zeta}_{i,p}^{k},\boldsymbol{\zeta}_{i,d}^{k}) \label{15a} \\
		&{\mathcal L}_{\eta}(\boldsymbol{x}_i^{k},\boldsymbol{x}_{i'}^{k+1},\boldsymbol{\zeta}_{i,p}^{k},\boldsymbol{\zeta}_{i,d}^{k})\le{\mathcal L}_{\eta}(\boldsymbol{x}_i^{k},\boldsymbol{x}_{i'}^{k},\boldsymbol{\zeta}_{i,p}^{k},\boldsymbol{\zeta}_{i,d}^{k}). \label{15b}
	\end{align}
\end{subequations}
If the surrogate optimality condition is satisfied, the line search on the sub-gradient direction forms acute angles towards optimal dual variables. Dual variables will be updated as:
\begin{subequations}
\begin{align}
	&\boldsymbol{\zeta}_{i,\tau,p}^{k+1}=\boldsymbol{\zeta}_{i,\tau,p}^{k}+\xi^{k}(\varpi_{i,p}\boldsymbol{x}_{i,\tau}^{k+1}-\varpi_{i',p}\boldsymbol{x}_{i',\tau}^{k}) \label{16a} \\
	
	&\boldsymbol{\zeta}_{i,\tau,d}^{k+1}=\boldsymbol{\zeta}_{i,\tau,d}^{k}+\xi^{k}(\varpi_{i,d}\boldsymbol{x}_{i,\tau}^{k+1}-\varpi_{i',d}\boldsymbol{x}_{i',\tau}^{k}). \label{16b}
\end{align} 
\end{subequations}

The stepsize toward the surrogate direction should be appropriately decided to ensure the convergence of the problem. The Polyak step-size rule is leveraged following \cite{bragin2018scalable} as:
\begin{align}
	&\xi^{k}=\alpha_{k}\frac{\xi^{k-1}||\gamma_p^{k-1}||_2}{||\gamma_p^{k}||_2},  \label{17}
\end{align} 
where $\gamma_p^{k}$ represents the primal residual at iteration $k$:
\begin{align}
	&\gamma_p^{k}=\sum_{i \in {\mathcal N}_p, i' \in {\mathcal D}/{\mathcal N}_p}\sum_{\tau \in {\mathcal T}}(\varpi_{i,p}\boldsymbol{x}_{i,\tau}^{k}-\varpi_{i',p}\boldsymbol{x}_{i',\tau}^{k}) \notag\\
	&+\sum\limits_{\substack{i \in {\mathcal N}_{iot,(fdc)}\\, i' \in {\mathcal N}_{fdc,(cdc)}}}\sum_{\tau \in {\mathcal T}} \boldsymbol(\varpi_{i,d}\boldsymbol{x}_{i,\tau}^{k}-\varpi_{i',d}\boldsymbol{x}_{i',\tau}^{k}),  \label{18}
\end{align} 
where the ancillary parameter $\alpha_{k}$ is updated as follow:
\begin{subequations}
\begin{align}
	&\alpha_{k}=1-\frac{1}{ck^{\theta}}, \quad \theta=1-\frac{1}{k^r}  \label{19a}\\
	 &c\ge 1, \quad 0 <r<1.  \label{19b}
\end{align} 
\end{subequations}
The penalty term is updated adaptively following:
\begin{subequations}
\begin{align}
&\boldsymbol{\eta}_{p(d)}^{k+1}=\boldsymbol{\eta}_{p(d)}^{k}/w, \text{ \eqref{15a} or \eqref{15a} is not satisfied}  \label{20a}\\
&\boldsymbol{\eta}_{p(d)}^{k+1}=\boldsymbol{\eta}_{p(d)}^{k} \cdot w, \text{ Otherwise}. \label{20b}
\end{align} 
\end{subequations}
Define the dual residual at iteration $k$ as:
\begin{align}
	&\gamma_d^{k}=\sum_{ i' \in {\mathcal D}/{\mathcal N}_p}\sum_{\tau \in {\mathcal T}}(\varpi_{i,p}\boldsymbol{x}_{i',\tau}^{k}-\varpi_{i',p}\boldsymbol{x}_{i',\tau}^{k-1}) \notag\\
	&+\sum_{i' \in {\mathcal N}_{fdc,cdc}}\sum_{\tau \in {\mathcal T}} \boldsymbol(\varpi_{i,d}\boldsymbol{x}_{i',\tau}^{k}-\varpi_{i',d}\boldsymbol{x}_{i',\tau}^{k-1}), k>0.  \label{21}
\end{align} 

\begin{figure}[!hbtp]
	{
	\centering
	\includegraphics[width=1.0\linewidth]{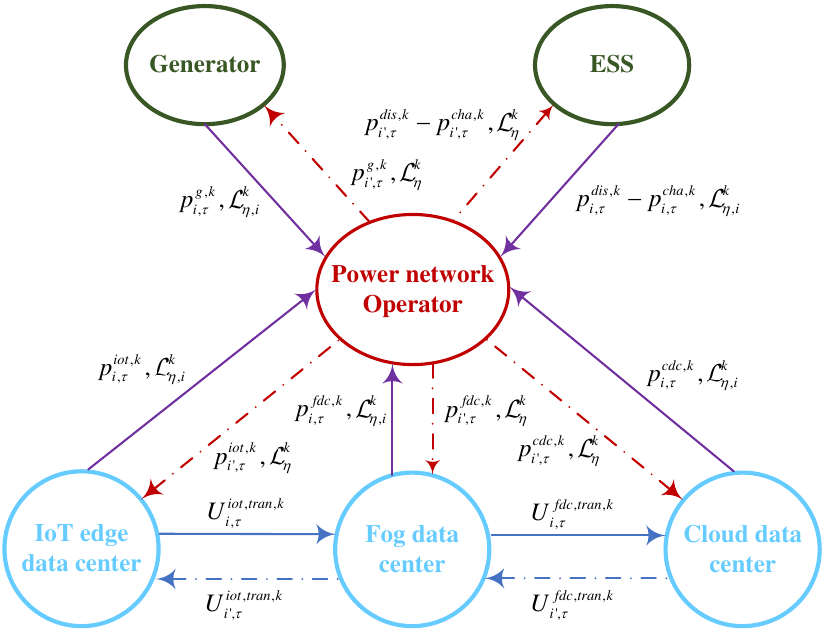}
	\caption{Variable sharing in distributed privacy-preserving computing.}
	\label{fig3.1}
	\vspace{-0.5cm}}
\end{figure}

%As illustrated in Fig. \ref{fig3.1}, 
At each iteration $k$, each agent in the system is responsible to send and receive data with other agents as required in Algorithm 1. {Specifically, as illustrated in Fig. \ref{fig3.1}, IoT data center, fog data center, and cloud data center need to interact with each other by sharing variables sequentially to keep communication conservation, i.e., the amount of incoming tasks should equal the amount of processed, transmitted, and stored tasks. These shared variables include $U_{i/i',\tau}^{iot,tran,k}$ and $U_{i/i',\tau}^{fdc,tran,k}$. Meanwhile, generator, ESS, allied with edge-fog-could data centers, must coordinate with the power network operator to keep energy conservation by exchanging active power consumption variables, i.e., $p_{i/i',\tau}^{g,k},p_{i/i',\tau}^{iot,k},p_{i/i',\tau}^{fdc,k},p_{i/i',\tau}^{cdc,k},p_{i/i',\tau}^{dis,k}-p_{i/i',\tau}^{cha,k}$. Moreover, to ensure the convergence, each agent needs to interact with the power network operator by sharing personal augmented Lagrangian function value, i.e., ${\mathcal L}_{\eta, i}^{k}$, and receiving the overall Lagrangian function value, i.e., ${\mathcal L}_{\eta}^{k}$, for deciding the satisfaction of surrogate Lagrangian conditions.} If the primal and dual residuals are below the stopping criteria, the distributed algorithm will be terminated. {We analyze the asymptotic complexity for the proposed distributed privacy-preserving algorithm using the $\mathcal{O}$ notation. Suppose there are total $N$ agents in the system. In each iteration, each agent transmits its own local variables from the $d-$dimensional set, i.e., $U_{i,\tau}^{iot,tran,k}$, $U_{i,\tau}^{fdc,tran,k}, p_{i,\tau}^{g,k},p_{i,\tau}^{iot,k},p_{i,\tau}^{fdc,k},p_{i,\tau}^{cdc,k}, p_{i,\tau}^{dis,k}-p_{i,\tau}^{cha,k}$ and $ {\mathcal L}_{\eta, i}^{k}$, to the power network operator or to its neighbors, which may include IoT edge centers, fog data centers, or cloud data centers. In return, the power network operator or its neighbors send back the updated shared or global variables from the $d-$dimensional set, i.e., $U_{i',\tau}^{iot,tran,k}$, $U_{i',\tau}^{fdc,tran,k}, p_{i',\tau}^{g,k},p_{i',\tau}^{iot,k},p_{i',\tau}^{fdc,k},p_{i',\tau}^{cdc,k}, p_{i',\tau}^{dis,k}-p_{i',\tau}^{cha,k}$ and $ {\mathcal L}_{\eta}^{k}$. In the uplink phase, each agent sends at most a $d-$dimensional vector. In  the downlink phase, each agent receives at most a $d-$dimensional vector. Therefore, communication cost per iteration per agent is $\mathcal{O}(d)$. Across $N$ agents, the communication cost per iteration for the system is $\mathcal{O}(Nd)$. Assuming the algorithm takes $K$ iterations to converge, the total asymptotic communication complexity is $\mathcal{O}(NdK)$. The asymptotic communication complexity depends on the dimension of the shared variables, the number of agents, and the number of iterations required for convergence.}

\subsection{Convergence Analysis}

{\bf Proposition 1:} With parameter settings in \eqref{19a}-\eqref{19b} and \eqref{20a}-\eqref{20b}, Algorithm 1 can ensure that the $\ell_1-$Lagrangian relaxed problem converges to optimal solutions of $\mathcal{P}_{\vartheta}$, and ultimately epi-converges to optimal solutions of $\mathcal{P}$.

\textit{Proof:} When surrogate optimality conditions, i.e, \eqref{15a} and \eqref{15b}, are met, the $\ell_1-$surrogate Lagrangian relaxed problem can converge to the optimal solutions \cite{bragin2015convergence}. Conversely, when surrogate optimality conditions are not met over iterations, penalty factors $\boldsymbol{\eta}_{p(d)}^{k}$ decrease and finally tend to zero. Convergence for the optimization problem under this situation can be theoretically guaranteed with the parameter settings in \eqref{19a}-\eqref{19b}, and \eqref{20a}-\eqref{20b} following \cite{bragin2018scalable}. {Meanwhile, the $\ell_1-$surrogate Lagrangian method can converge to optimal solutions with a linear convergence rate. From \cite{bragin2015convergence,nedic2001convergence}, the relation between dual variables and surrogate Lagrangian function value can be constructed as: 
\begin{align}
	& \beta \cdot||\boldsymbol{\eta}_{p(d)}^{*}-\boldsymbol{\eta}_{p(d)}^{k}||^2 \notag\\
	&\le {\mathcal L}_{\eta}^* (\boldsymbol{x}_i,\boldsymbol{x}_{i'},\boldsymbol{\zeta}_{i,p},\boldsymbol{\zeta}_{i,d})-{\mathcal L}_{\eta} (\boldsymbol{x}_i,\boldsymbol{x}_{i'},\boldsymbol{\zeta}_{i,p},\boldsymbol{\zeta}_{i,d}), \label{19c}
\end{align} 
where $\boldsymbol{\eta}_{p(d)}^{*}$ and ${\mathcal L}_{\eta}^* (\boldsymbol{x}_i,\boldsymbol{x}_{i'},\boldsymbol{\zeta}_{i,p},\boldsymbol{\zeta}_{i,d})$ refer to the optimal dual variables and Lagrangian function value, respectively. Thus, the parameter $\beta$ exists, which satisfies the condition that $0<\beta<\frac{1}{2\xi^{k}}$. The distance between dual variables and their optimal values is proportional to the violation of global constraints. Dual variables $\boldsymbol{\eta}_{p(d)}^{k}$ can convergence the optimal values with an approximate linear rate of $\sqrt{1-2\beta\xi^{k}}<1$ outside a sphere centered at $\boldsymbol{\eta}_{p(d)}^{*}$. Recall Lemma 2, the optimality condition in \eqref{12} holds true when the parameter $\vartheta$ is made arbitrarily small. This implies that the optimal solutions of the relaxed MINLP problem epi-converge to those of the original MIQCQP problem. Overall, by leveraging the surrogate optimality conditions and adaptively adjusted stepsizes, the proposed approach is theoretically guaranteed to yield optimal solutions the original problem $\mathcal{P}$ as the parameter $\vartheta$ is sufficiently small. Furthermore, we verify the effectiveness and convergence of the proposed approach via the numerical simulations.}

\section{Numerical Simulations}
\subsection{Simulation Setup}
The code is implemented in Matlab 2020 on a computer with 6-core Intel i5-10500 CPU@3.10GH, and the state-of-the-art commercial solver, Gurobi 11.0, is employed to solve the centralized problem and decomposed sub-problems. {The standard IEEE 15-bus, 33-bus, and 85-bus distribution systems \cite{zimmerman2010matpower} penetrated with data centers are adopted to verify the optimality and efficiency of the proposed distributed privacy-preserving approach. The day-ahead real PV and wind power generation data are derived from \cite{jenkins2019hybrid}, and the communication traffic data are adopted from \cite{gao2023task}. Day-ahead locational marginal prices are derived from the UK electricity market \cite{Nordpool2023new}. The load task requests at the IoT data center $j$, i.e., $S_{j,\tau}^{iot}$, are randomly generated from the Poisson distribution \cite{tong2022dynamic}. These data used in this paper are available in the public repository \cite{junhong2024}. The real-time time-slot is set to be the same as dispatch time-slot for simplicity in this paper. Parameters for updating step size are set as: $c=200$, $r=\frac{1}{\sqrt{k}}$, and $w=1.01$.}

\subsection{IEEE 15-Bus Systems with Simplified Data Centers}
\begin{figure}[!hbtp]
	\centering
	\includegraphics[width=1.0\linewidth]{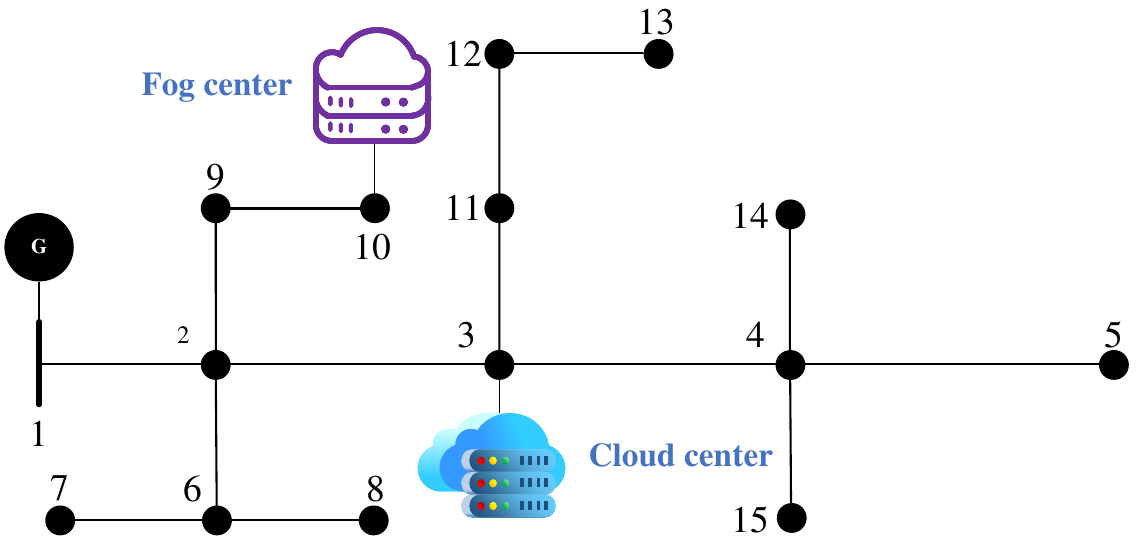}
	\caption{IEEE 15-bus system with simplified data centers.}
	%\vspace{-0.5cm}
	\label{fig3}
\end{figure}

{The synergy of hierarchical data centers and power networks constitutes a non-convex and non-smooth optimization problem, posing significant challenges in obtaining near-optimal solutions, even with centralized methods. To validate the near-optimality of the proposed distributed privacy-preserving approach, a case study is thus designed using the simplified data center structure with lower complexity. Specifically, The optimality and convergence of the proposed distributed approach are first verified on the IEEE 15-bus systems with simplified data centers.} The stopping criteria for primal and dual residuals are set as $\gamma_p^k \le 10^{-2}$ and  $\gamma_d^k \le 10^{-2}$ or $|\gamma_p^{k+1}-\gamma_p^k| \le 10^{-6}$. The initial step size is set to 1.3. As illustrated in Fig. \ref{fig3}, the traditional diesel generator is positioned at bus 1, with a fog data center located at bus 10. Additionally, a cloud data center is placed at bus 3 together with the combined renewable energy generators (wind and power generators). The effectiveness of RNMDT method is verified under both the centralized and distributed settings with different segment parameter $\vartheta$. When $\vartheta=0$, the RNMDT method becomes a set of classical McCormick envelopes \cite{mitsos2009mccormick}, which represents the loosest approximation. As denoted in Table \ref{tab1}, the original optimization problem contains 216 binary variables, with additional 48 quadratic terms in the objective function (48o) and 336 quadratic terms in constraints (336c). The original optimization problem is a non-convex MIQCQP. Through approximation by RNMDT with slightly increased binary variables, the quadratic terms in the objective function are removed, left with quadratic constraints, i.e., convex second-order cone constraints denoted by \eqref{1e}. As in Table \ref{tab2}, the classical McCormick envelope method, i.e., $\vartheta=0$, obtains a lower cost than the optimal solution. As the segment parameter decreases, solutions obtained by the RNMDT tend to the optimal one. Meanwhile, the relaxed problem by the RNMDT requires much lower computational costs than the original problem.

\begin{figure}[hbtp]
	\centering
	\includegraphics[width=1.0\linewidth]{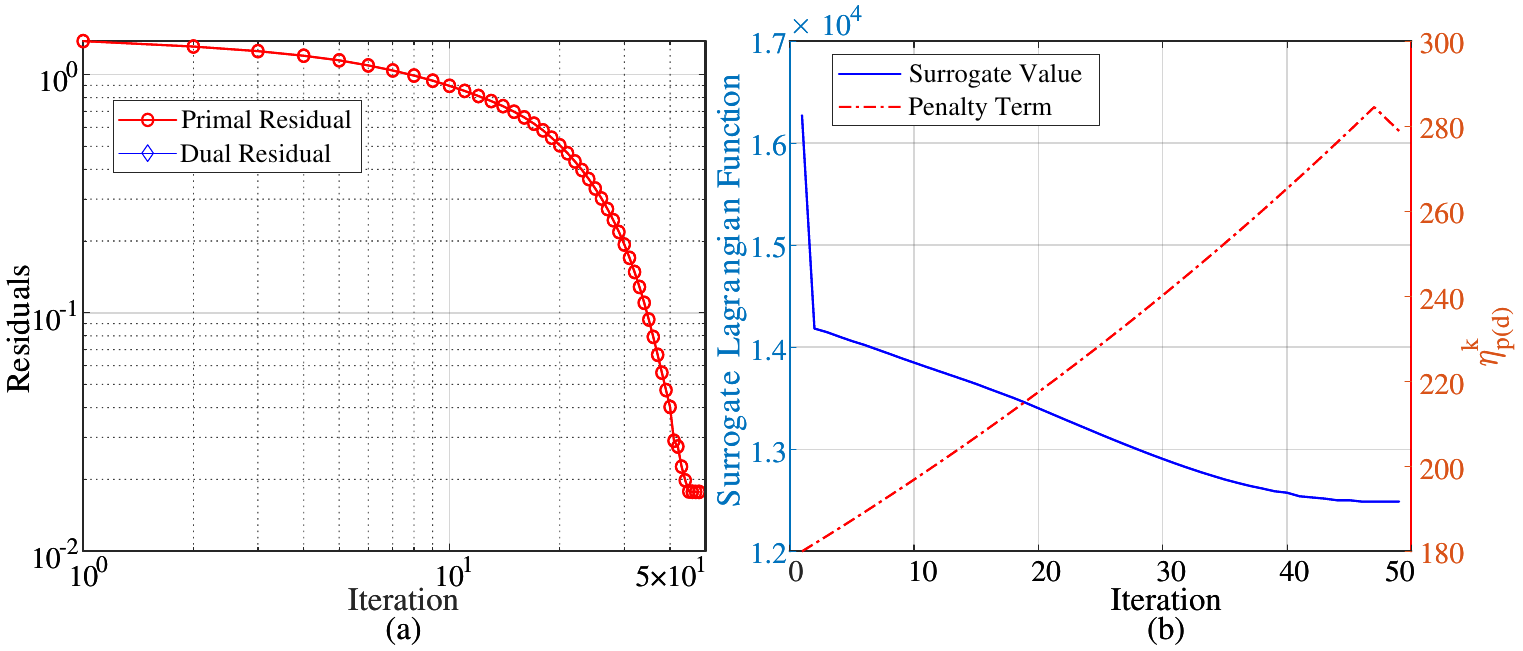}
	\caption{Convergence curves. (a) primal and dual residuals. (b) evolution of the surrogate value and penalty term.}
%\vspace{-0.5cm}
	\label{fig4}
\end{figure}

\begin{figure}[!hbtp]
	{
		\centering
		\includegraphics[width=0.8\linewidth]{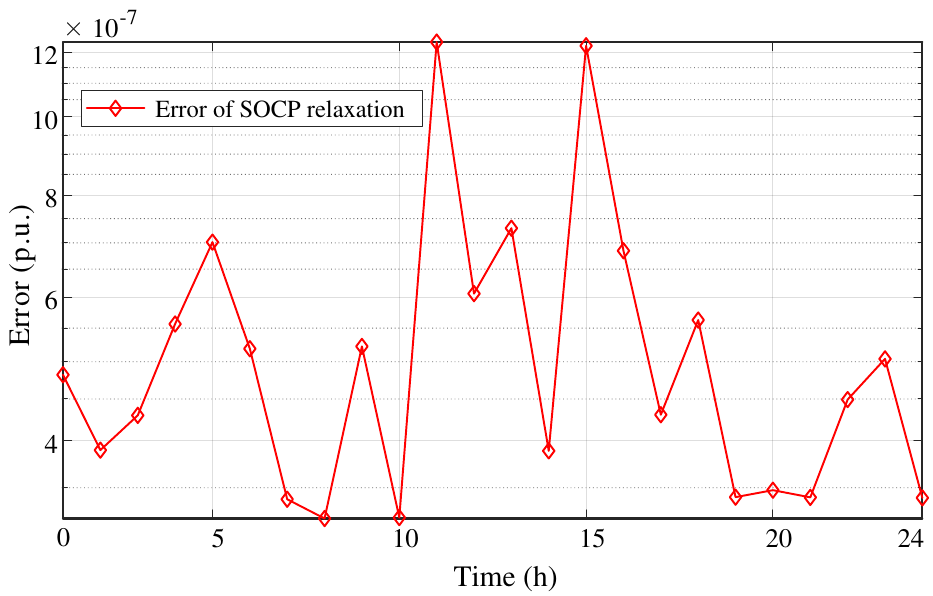}
		\caption{Error for the SOCP relaxation.}
		%\vspace{-0.5cm}
		\label{fig_socp}
	}
\end{figure}

As illustrated in Fig. \ref{fig4} (a), the primal and dual residuals drop to the stopping criteria around 49 iterations when $\vartheta=-3$. As in Fig. \ref{fig4} (b), the adaptive mechanism for penalty term contributes to the further decreasement of surrogate Lagrangian function value. With the decrement of segment parameter under the distributed setting, solutions obtained by the RNMDT tend to the optimal one as in Table \ref{tab3}. Meanwhile, the original problem solved directly in the distributed manner obtains the solution with higher gaps due to its inherent high non-convexities introduced by the quadratic terms. These results illustrate the exactness and effectiveness of relaxing quadratic terms through RNMDT. {Moreover, the exactness of the SOCP relaxation is verified. At each dispatch time slot, the largest errors, i.e., the error from equation \eqref{1e}, for all buses are obtained. As illustrated in the Fig. \ref{fig_socp}, the relaxation is shown to be exact, with the largest errors across all buses at each dispatch time slot being negligible.}

%variables for 15 bus system%
\begin{table}[htbp]
	\centering
	\caption{Statistics of Model Parameters for 15-Bus System}
	\label{tab1}
	\begin{tabular}{cccc}
		\toprule  
		& Binary Variables & Continuous Variables & Quadratic Terms\\ 
		\cmidrule(r){2-4}
		{Original }&216 &4896&48o+336c\\
		{0}&216 &2978&336c\\
		{-1}&217 &2989&336c\\
		{-3}&230 &3027&336c\\
		{-12}&504 &5856&336c\\
		\bottomrule 
	\end{tabular}
	\vspace{-0.5cm}
\end{table}\

%for the RNMT algorithm accuray%
\begin{table}[htbp]
	\centering
	\caption{Centralized Method using Different Parameters}
	\label{tab2}
	\begin{tabular}{cccccc}
		\toprule  
		&Original & 0 &-1 & -3 &-12\\ 
		\cmidrule(r){2-6}
		{Value}&12468.31&12462.92 &12468.31&12468.31&12468.31\\
		{Time (s)}&4.68 &3.51 &3.80&3.93&7.26\\
		\bottomrule 
	\end{tabular}
	\vspace{-0.5cm}
\end{table}\

%for the RNMT algorithm accuray%
\begin{table}[!htbp]
	\centering
	\caption{Distributed Method using Different Parameters}
	\label{tab3}
	\begin{tabular}{cccccc}
		\toprule  
		&Original & 0 &-1 & -3 &-12\\ 
		\cmidrule(r){2-6}
		{Value}&12465.56&12460.72&12466.33&12467.62&12467.61\\
		{Time (s)}&267.40&252.65&253.07&273.03&288.24\\
		{Iteration}&47&47&46&49&50\\
		\bottomrule 
	\end{tabular}
	\vspace{-0.5cm}
\end{table}\

{To further demonstrate the optimality of the proposed distributed approach (M3), we further compare it with the centralized method (M1) and the classic ADMM (M2) \cite{boyd2011distributed}. Define the relative absolute error as $|O_{M1}-O_{M2/M3}|/O_{M1}\cdot 100\%$, where $O_{M1}$, $O_{M2}$, and $O_{M3}$ denote the optimal objective values obtained by M1, M2 and M3, respectively. For the distributed approach, the whole optimization problem is decomposed into a set of sub-problems for each agent, i.e., the power network operator, fog data center (FDC), cloud data center (CDC), and power plant generator. Each agent solves its own decomposed sub-problem and simultaneously shares limited data with other agents, which preserves privacy of personal data for each agent. As in Table \ref{tab4}, when $\vartheta=-3$, relative absolute errors for all the sub-problems are below $1\%$. Meanwhile, M3 obtains a solution with the overall error of $0.006\%$ compared to the optimal solutions obtained from M1, which demonstrates that M3 can derive high-quality solutions for non-convex problem in a privacy-preserving manner. For the classic ADMM, i.e., M2 , all the parameter settings are kept the same as M3 except for the step-size. The step-size for M2 employed the initial step-size for M3, and we set the 200 iteration as the maximum limit. From the simulation results, M2 actually obtains a relatively large primal error, i.e., $\gamma_p^k > 10^{-1}$, within the maximum limits. For M2, the total relative error with regarding to the optimal objective function derived by M1 is $0.367\%$, with the relative error of $9.41\%$ for the power network and the relative error of $7.69\%$ for the generator. These simulation results demonstrate the superiority of the proposed distributed approach, i.e., M3, over the traditional distributed optimization method, i.e., M2.}

\begin{table}[htbp]
	{
	\centering
	\caption{Costs of Different Algorithms (EUR)}
	\label{tab4}
	\scriptsize
	\begin{tabular}{cccccc}
		\toprule  
		&FDC &CDC & Generator &Power network & Total \\ 
		\cmidrule(r){2-6}
		{M1}&11023.94 &815.42&614.07&14.88 & {\bf12468.31}\\
		{M2}&11030.34 &811.85&566.85&13.48 & {\bf12422.52}\\
		{M3}&11030.34 &811.80&610.63&14.85 & {\bf12467.62}\\	
    	{Error (M2)}&0.06\% &0.44\%&7.69\%&9.41\% & {\bf 0.367\%}\\
    	{Error (M3)}&0.06\% &0.44\%&0.56\%&0.20\% & {\bf 0.006\%}\\
		\bottomrule 
	\end{tabular}}
	\vspace{-0.5cm}
\end{table}\

\begin{figure}[!hbtp]
	{
		\centering
		\includegraphics[width=0.82\linewidth]{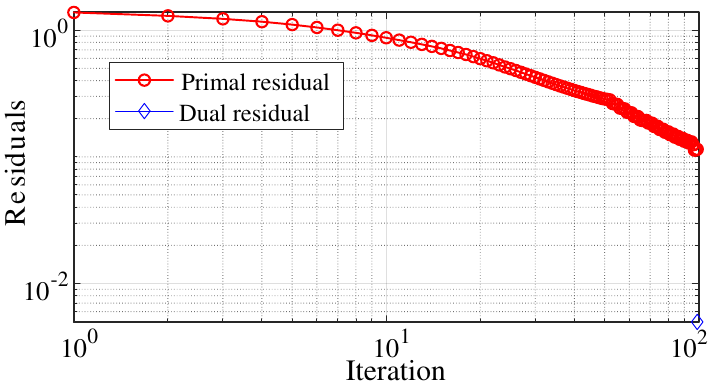}
		\caption{Convergence curve for ADMM.}
		%\vspace{-0.5cm}
		\label{fig5.2}}
\end{figure}

\begin{figure}[!hbtp]
	{
	\centering
	\includegraphics[width=0.82\linewidth]{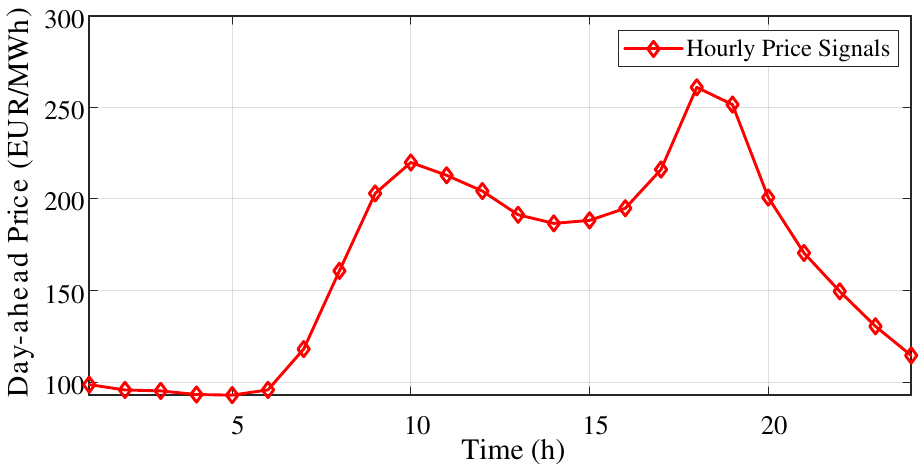}
	\caption{Day-ahead price signals.}
	\vspace{-0.5cm}
	\label{fig5.1}}
\end{figure}

\begin{figure}[!hbtp]
\centering
\includegraphics[width=0.8\linewidth]{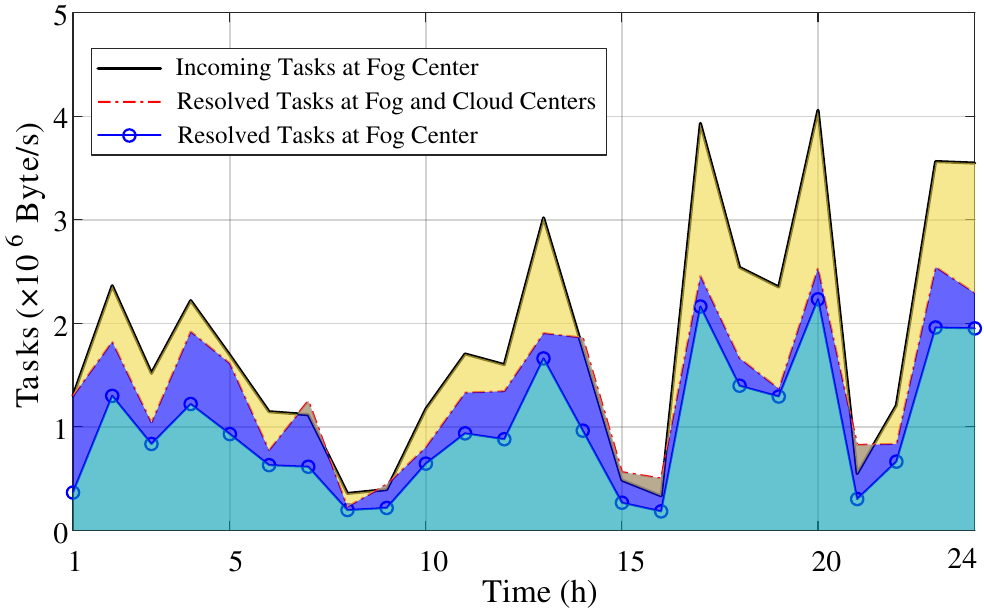}
\caption{Task allocations between fog and cloud centers.}
%\vspace{-0.5cm}
\label{fig5}
\end{figure}

\begin{figure}[hbtp]
	\centering
	\includegraphics[width=1.0\linewidth]{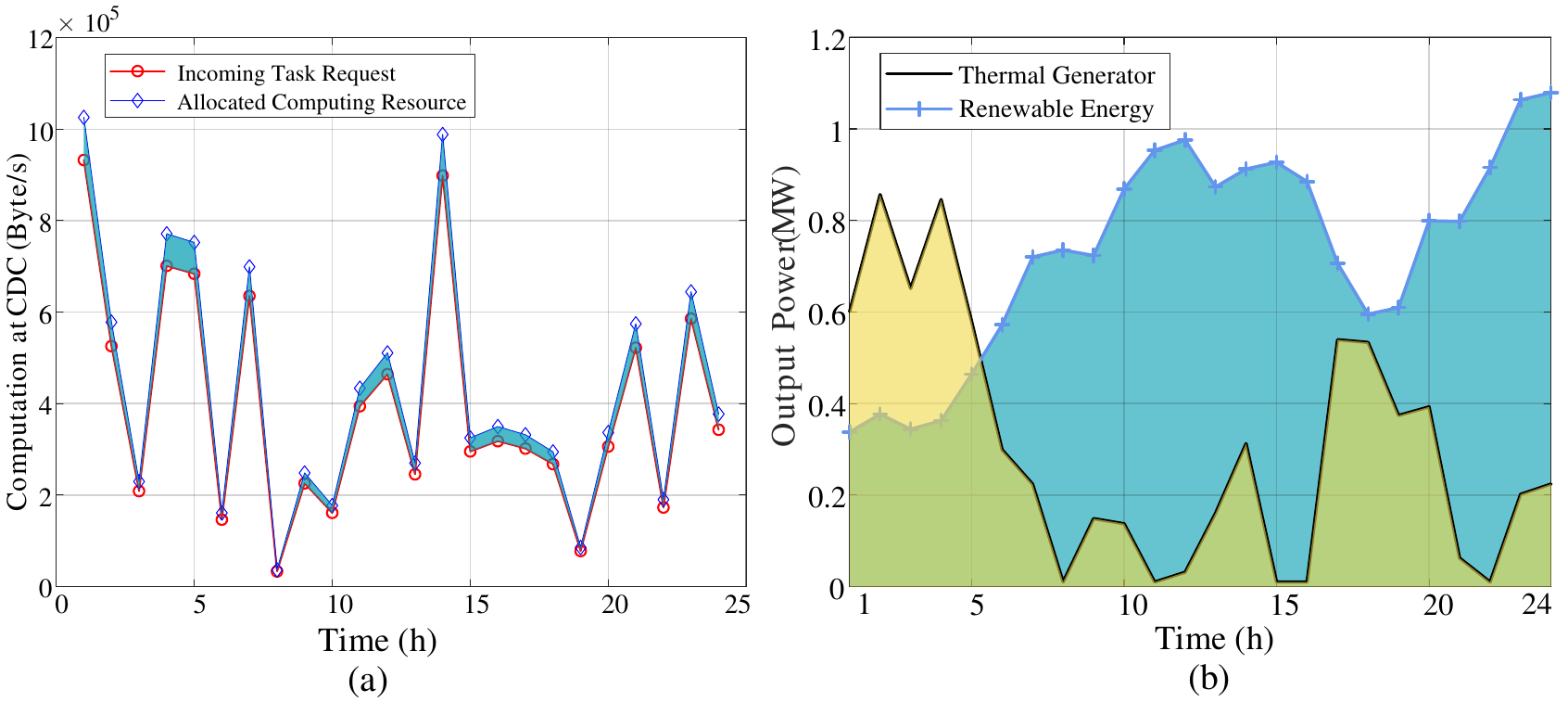}
	\caption{Evolution of task demands and energy generations. (a) resource allocation at cloud data center (b) power supplies}
	%\vspace{-0.5cm}
	\label{fig6}
\end{figure}

As in Fig. \ref{fig5}, incoming tasks at the fog data center are allocated to be either resolved locally at the fog data center, or stored in the data queue, or transferred remotely to cloud data center for computations to reach the minimum operational cost for the integrated systems. As illustrated in Fig. \ref{fig6} (a), to meet the quality of services, the cloud data center should reserve adequate computing resources for the random incoming task requests from fog data center. Meanwhile, the cloud data center will priorly utilize the local renewable energy generations to perform computations for the incoming requests, and the diesel generator will coordinate with uncertain renewable energy generations to supply stable electricity for the whole system as illustrated in Fig. \ref{fig6} (b). {Specifically, The computational load is optimally scheduled based on the incoming task requests, hourly price signals, volume of the data queues, and the availability of renewable energy generation with zero marginal costs. As shown in Fig. \ref{fig6} (b), renewable generation from the 1st to 5th hours is relatively low compared to other time periods. However, as illustrated in Fig. \ref{fig5.1}, the price signals during these hours are low, and the incoming tasks at the fog data centers are also low as depicted by the solid black line in Fig. \ref{fig5}. In this scenario, the traditional thermal generator is scheduled to supply sufficient power due to the low electricity prices, allowing a large portion of the computational load to be processed by the fog and data centers. Meanwhile, during the 10th and 20th hours, with high price signals and abundant renewable energy generation, the traditional thermal generator is less scheduled to supply power, resulting in a balanced overall power supply. Despite the relatively high volume of incoming task requests during these hours, a portion of the requests is stored in the data queue for later processing, balancing the computational load effectively.}

%centralized hard
%distriuted ok
%privacy-preserving
%performance

\subsection{IEEE 33-Bus Systems with Hierarchical Data Centers}
\begin{figure}[!hbtp]
	\centering
	\includegraphics[width=1.0\linewidth]{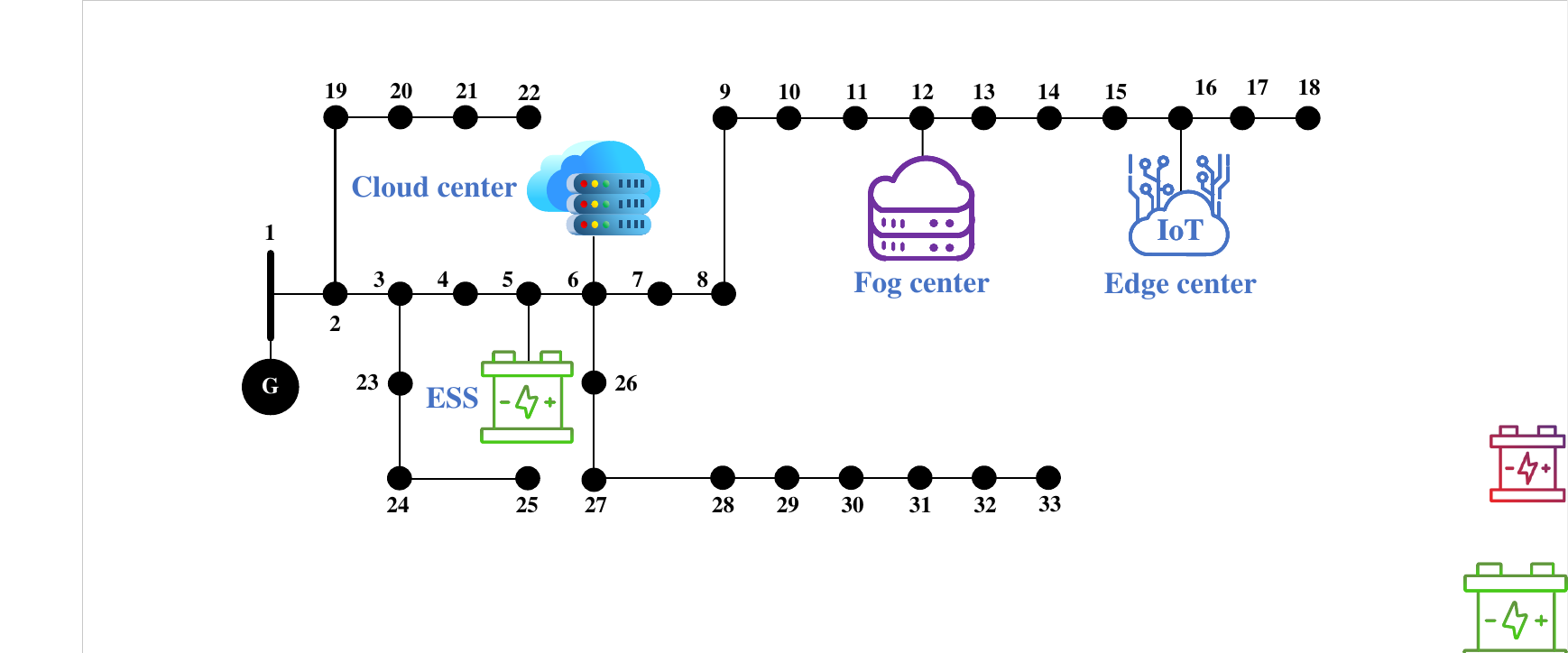}
	\caption{IEEE 33-bus system with hierarchical data centers.}
	\vspace{-0.5cm}
	\label{fig7}
\end{figure}

The effectiveness and scalability of the proposed distributed approach are further verified on the IEEE 33-bus systems with hierarchical data centers. As illustrated in Fig. \ref{fig7}, the energy storage system (ESS) is deployed at bus 5. An edge data center is positioned at bus 16, together with the fog data center at bus 12 and cloud data center at bus 6, forming the three-tier hierarchical data centers. Due to the existence of additional ESS and edge data center, number of quadratic terms in the objective function and integer variables increases as given in Table \ref{tab5}, complicating the optimization problem. The stopping criteria for primal and dual residuals are set as $\gamma_p^k \le 10^{-6}$ and $\gamma_d^k \le 10^{-6}$ or $|\gamma_p^{k+1}-\gamma_p^k| \le 10^{-6}$ for better performance of the distributed approach, and the initial step size is set to 1.3.

%variables for 33 bus system%
\begin{table}[htbp]
	\centering
	\caption{Statistics of Model Parameters for 33-Bus System}
	\label{tab5}
	\begin{tabular}{cccc}
		\toprule  
		& Binary Variables & Continuous Variables & Quadratic Terms\\ 
		\cmidrule(r){2-4}
		{Original }&240 &12217&144o+768c\\
		{0}&240 &6470&768c\\
		{-5}&456 &7043&768c\\
		{-8}&594 &7388&768c\\
		{-12}&785 &7866&768c\\
		\bottomrule 
	\end{tabular}
	\vspace{-0.5cm}
\end{table}\

\begin{table}[htbp]
	\centering
	\caption{Centralized Method using Different Parameters}
	\label{tab6} %location error
	\begin{tabular}{cccccc}
		\toprule  
		& Original & 0 & -5 & -8& -12 \\ 
		\cmidrule(r){2-6}
		{Value}&-2575.43& -15243.73 &-2595.88&-2789.21&-2509.52\\
		{Gap}&144.77\% &0.00\% &72.12\%&76.10\%&63.14\% \\
		{Time (s)}&36000.00 &48.85&36000.00&36000.00&36000.00\\
		\bottomrule 
	\end{tabular}
	\vspace{-0.5cm}
\end{table}\

Since the objective function contains the Lyapunov drift terms, the combined equivalent operational cost can be negative. As denoted in Table \ref{tab6}, when $\vartheta=0$, the classical McCormick envelope method obtains an optimal solution within $48.85 s$. However, the obtained value is far from the global optimal value due to the highly loose relaxation by the classical McCormick envelope method. Without relaxation by RNMDT, the centralized manner cannot derive the optimal solution within finite time, i.e., 10 hours, and the optimality gap is nearly 144.78\%. RNMDT can facilitate the derivation of optimal solutions, while the optimality gap for all cases still remains high (over 63\%), which is unacceptable for solving the day-ahead dispatch problem. Simulation results demonstrate that the traditional centralized method combined with efficient relaxations still cannot be leveraged to derive high-quality solutions for the complex synergy problem of hierarchical data centers penetrated power network. To deal with this challenge and meanwhile preserve privacy for agents in the integrated system, the efficiency of distributed $\ell_1-$surrogate Lagrangian method combined with RNMDT is further verified.

\begin{table}[htbp]
	\centering
	\caption{Distributed Method using Different Parameters}
	\label{tab7}
	\begin{tabular}{cccccc}
		\toprule  
		& Original & 0 & -5 & -8& -12 \\ 
		\cmidrule(r){2-6}
		{Value}&-&-15103.24 &-2515.52&-2433.30&-2433.47\\
		{Iteration}&- &149&162&182&170\\
		{Time$_1$ (s)}&- &1784.96&4935.27&5983.04&6669.85\\
  		{Time$_2$ (s)}&- &297.49&822.55&997.17&1111.64\\
		\bottomrule 
	\end{tabular}
	\vspace{-0.5cm}
\end{table}\

\begin{figure}[hbtp]
	\centering
	\includegraphics[width=1.01\linewidth]{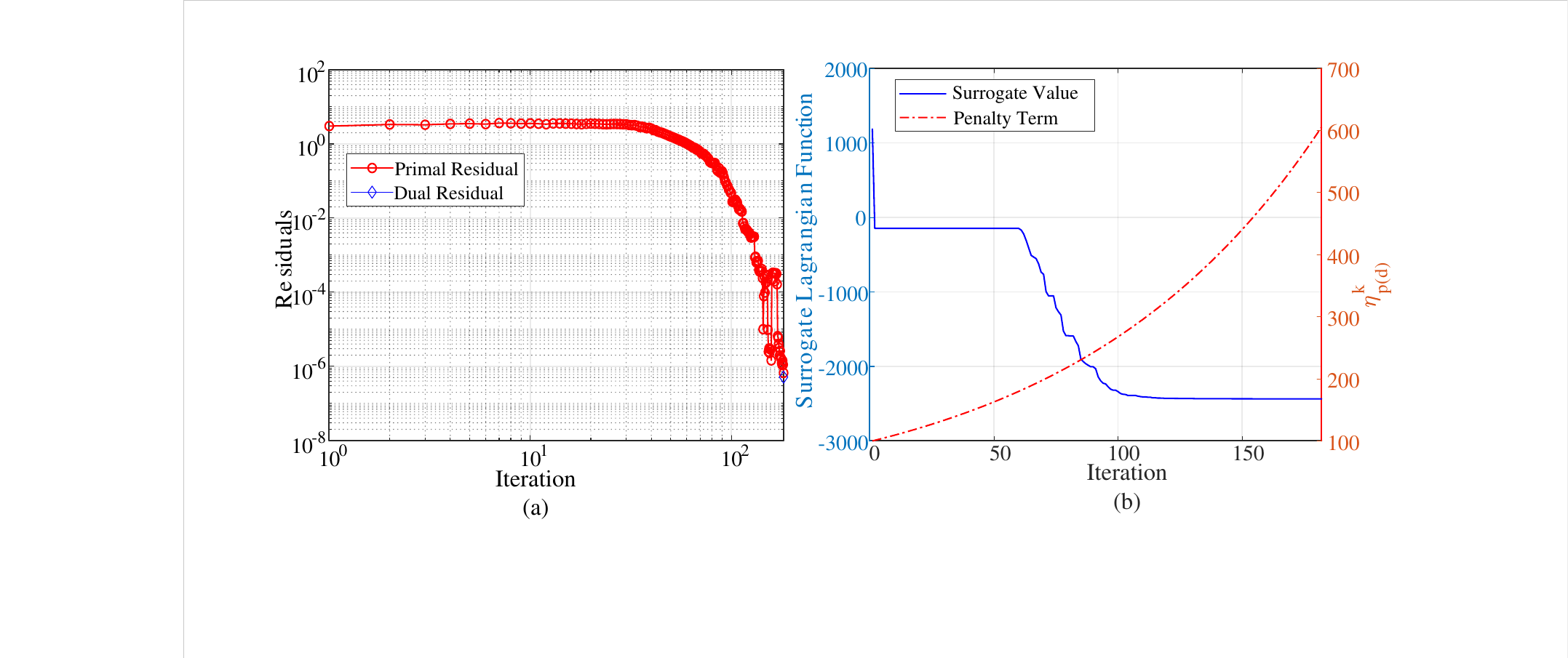}
	\caption{Convergence curves. (a) primal and dual residuals. (b) evolution of the surrogate value and penalty term.}
	%\vspace{-0.5cm}
	\label{fig8}
\end{figure}

\begin{figure}[hbtp]
	\centering
	\includegraphics[width=0.8\linewidth]{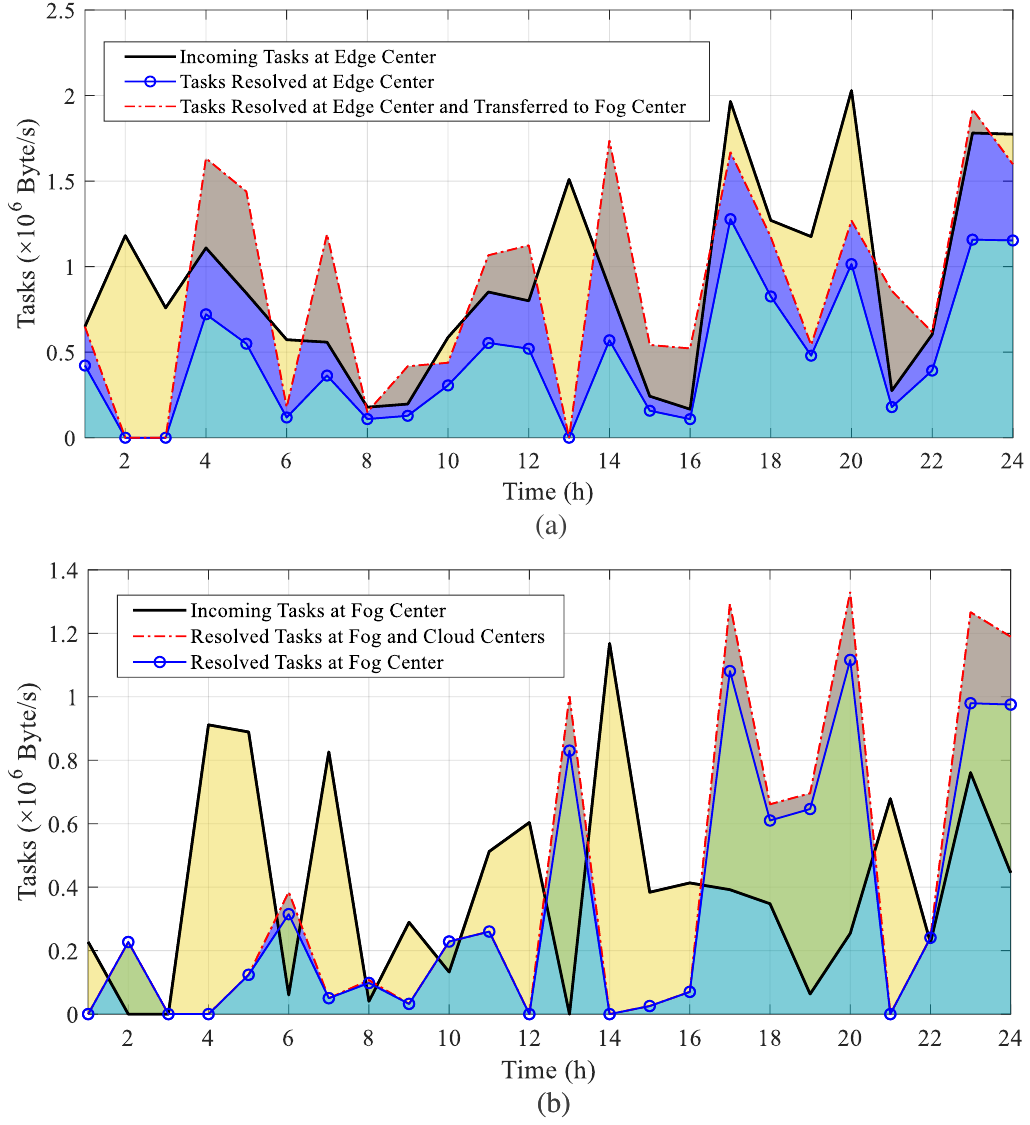}
	\caption{Task allocations. (a) between edge and fog centers. (b) between fog and cloud centers.}
	%\vspace{-0.5cm}
	\label{fig9}
\end{figure}

For the distributed setting, the whole optimization problem is decomposed into sub-problems for each agent, i.e., the power network operator, IoT edge data center (EDC), energy storage system (ESS), FDC, CDC, and power plant generator. As denoted in Table \ref{tab7}, the proposed distributed privacy-preserving approach is verified with different segment parameter $\vartheta$. Time$_1$ refers to the overall total time costs for deriving solutions for all agents, while Time$_2$ refers to the ideal agent-based average time in the distributed computation scenario. However, the surrogate Lagrangian method cannot work for the original optimization problem (without relaxation for the quadratic terms). With the decrement of parameter $\vartheta$, the distributed algorithm can converge to the stable near-optimal solutions. Specifically, when $\vartheta=-8$, the distributed algorithm can converge to the required criteria within 182 iterations and $5984s$ (ideally 16.62 minutes for each agent in the distributed setting). As illustrated in Fig. \ref{fig8} (a) and \ref{fig8} (b), the surrogate optimality condition for this case is always satisfied and the penalty factors are incremented, contributing to the convergence of distributed algorithm. Dispatch results for the three-tier hierarchical data centers are exhibited. As illustrated in Fig. \ref{fig9} (a), incoming tasks at EDC are mostly resolved locally, which is denoted by the area enclosed by the solid blue circled line and x-axis. Beyond being stored in the data queue, partial tasks are also transferred to the fog data center for processing, as denoted by the area enclosed by the dash read line and solid blue circled line. For fog data center, as illustrated in Fig. \ref{fig9} (b) , incoming tasks from the EDC are also optimally allocated (to be either processed locally, stored in the data queue, or transferred to the cloud data center), to reach the optimal energy efficiency for the whole system.

{
\subsection{IEEE 85-Bus Systems with Hierarchical Data Centers}
The effectiveness and scalability of the proposed distributed approach are also verified on the large-scale IEEE 85-bus systems with hierarchical data centers. The power network topology for the IEEE 85-bus system is available in \cite{zimmerman2010matpower}, where the location of each entity is given in Table \ref{tab8}. Compared to the 15-bus and 33-bus cases with the same settings, this case contains more continuous and binary variables, as well as more quadratic terms in the constraints as illustrated in Table \ref{tab9}. For this case, the stopping criteria for primal and dual residuals are set the same as the 15-bus case, i.e., $\gamma_p^k \le 10^{-2}$ and  $\gamma_d^k \le 10^{-2}$ or $|\gamma_p^{k+1}-\gamma_p^k| \le 10^{-6}$. The initial step size is set to 1.3.
\begin{table}[htbp]
	{
		\centering
		\caption{Location of entities in the 85-bus system}
		\label{tab8}
		\scriptsize
		\begin{tabular}{ccccccc}
			\toprule  
			& Thermal generator & Renewable & Edge center &FDC &CDC & ESS \\ 
			\cmidrule(r){2-7}
			{Bus}&1 &50&74 &40&50&15 \\
			\bottomrule 
	\end{tabular}}
	\vspace{-0.5cm}
\end{table}\

\begin{table}[htbp]
	\centering
	{
	\caption{Statistics of Model Parameters for 85-Bus System}
	\label{tab9}
	\begin{tabular}{cccc}
		\toprule  
		& Binary Variables & Continuous Variables & Quadratic Terms\\ 
		\cmidrule(r){2-4}
		{Original }&240 &30937&144o+2016c\\
		{0}&240 &31129&2016c\\
		{-1}&288 &31273&2016c\\
		{-3}&384 &31561&2016c\\
		{-5}&480 &31849&2016c\\
		{-8}&624 &32281&2016c\\
		\bottomrule 
	\end{tabular}
	}
	%\vspace{-1.0cm}
\end{table}\

The original synergy problem for the 85-bus system with hierarchical data centers can not solved by the centralized method and also can not be directly solved by the distributed optimization method within the finite time. Through the reformulation by RNMDT, the near-optimal solutions can be derived by the proposed customized $\ell_1-$surrogate Lagrangian method. Simulation results in Table \ref{tab10} indicate that the proposed distributed approach can derive acceptable results within an ideal agent-based average time of $3459.10s$, which demonstrates the effectiveness and scalability of the proposed distributed approach in solving the large-scale, non-convex, and non-smooth synergy problems. These simulations are carried out on a personal computer to demonstrate feasibility. In real applications, more powerful computing facilities can further reduce the computation time.
\begin{table}[htbp]
	\centering
	{
	\caption{Distributed Method using Different Parameters}
	\scriptsize
	\label{tab10}
	\begin{tabular}{ccccccc}
		\toprule  
		& 0 & -1 & -3 & -5 & -8\\ 
		\cmidrule(r){2-7}
		{Value}&-1321.60&-136.57&1297.26&1404.16&1465.45\\
		{Iteration}&208&208&209&211&215\\
		{Time$_{1}$ (s)}&806.23&11860.28&14624.55&17413.45&20754.60\\
		{Time$_{2}$ (s)}&134.37&1976.71&2437.43&2902.24&3459.10\\
		\bottomrule 
	\end{tabular}
	}
	\vspace{-0.5cm}
\end{table}\
}

%\begin{table}[htbp]
%	\centering
%	\label{tab6}
%	\caption{Centralized Method using Different Parameters}
%	\begin{tabular}{cccc}
%		\toprule  
%		& Total Value (EUR) & Gap & Time (s) \\ 
%		\cmidrule(r){2-4}
%		{Original}&-2575.43&144.77\%&36000.00\\
%		{0}&-15243.73 &0.00\% &48.85\\
%		{-3}&-2783.28&78.74\% &36000.00\\
%		{-5}&-2595.88 &72.12\%&36000.00\\
%		{-8}&-2789.21&76.10\% &36000.00 \\
%		{-12}&-2509.52&63.14\% &36000.00\\
%		\bottomrule 
%	\end{tabular}
%	%\vspace{-0.5cm}
%\end{table}\

%\begin{table}[htbp]
%	\centering
%	\label{tab7}
%	\caption{Distributed Method using Different Parameters}
%	\begin{tabular}{cccc}
%		\toprule  
%		& Total Value (EUR) & Iteration & Time (s) \\ 
%		\cmidrule(r){2-4}
%		{Original}&-&-&-\\
%		{0}&-15103.24&149 &1784.96\\
%		{-3}&-2743.58&166 &2461.66\\
%		{-5}&-2515.52&162 &4935.27\\
%		{-8}&-2433.30&182 &5983.04 \\
%		{-12}&-2433.47&170 &6669.85\\
%		\bottomrule 
%	\end{tabular}
%	%\vspace{-0.5cm}
%\end{table}\

\section{Conclusions}
To address privacy leakage and combinatorial explosion concerns in the highly non-convex synergy problem of hierarchical data center penetrated power networks, we propose a near-optimal privacy-preserving distributed approach. The normalized multi-parametric disaggregation technique is leveraged to reformulate the non-convex mixed integer quadratically constrained quadratic programming into a mixed integer non-linear programming with the arbitrary accuracy. To further overcome the non-smoothness of the mixed integer problem, the customized $\ell_1-$surrogate Lagrangian relaxation method with convergence guarantees is proposed to solve the problem in a distributed privacy-preserving manner. {Simulation results demonstrate that the proposed approach: i) derives high-quality near-optimal solutions and ensures convergence of the synergy problem; ii) preserves privacy of personal data for different agents in the integrated systems; iii) has high computational efficiency for the complicated synergy problem; iv) can leverage the flexible resource allocation capabilities of the hierarchical data center architecture, further facilitating peak load balancing in the power network.}
{Future directions include developing measures of optimality gap and control mechanisms to boost performance of the customized $\ell_1-$surrogate Lagrangian method for synergy problems of the general and realistic integrated power networks.}

%{\appendix
%	
%	\subsection{Proof of Lemma 1}
%     
%     .$\hfill\blacksquare$
%     
%   	\subsection{Proof of Lemma 2}
%     
%     .$\hfill\blacksquare$
%}

\bibliographystyle{IEEEtran}
\bibliography{index}

\end{document}